\documentclass[man]{apa7}
\usepackage[american]{babel}
\usepackage{fontspec}

\usepackage{amsmath}
\usepackage{enumitem}
\usepackage{tabularx}
\usepackage{multirow}
\usepackage{tikz}
\usetikzlibrary{shapes.geometric, arrows, positioning}
\usepackage{soul}
\usepackage{xspace}
\usepackage{multirow}
\usepackage{tabularx}
\usepackage{booktabs}
\usepackage{longtable}   

\usepackage{enumitem}
\usepackage{tabularray}
\usepackage{float}
\usepackage{graphicx}
\usepackage{codehigh}
\usepackage{pdflscape}

\usepackage[normalem]{ulem}
\UseTblrLibrary{booktabs}
\UseTblrLibrary{siunitx}

\NewTableCommand{\tinytableDefineColor}[3]{\definecolor{#1}{#2}{#3}}

\usepackage{caption}
\usepackage{subcaption}

\usepackage{csquotes}
\usepackage[style=apa,sortcites=true,sorting=nyt,backend=biber,natbib=true]{biblatex}
\DeclareLanguageMapping{american}{american-apa}
\title{Taming toxic talk: Using chatbots to intervene with users posting toxic comments}
\shorttitle{Taming Toxic Talk}

\leftheader{Foote et al.}

\authorsnames[1,2,1,3,1,4,1]{Jeremy Foote, Deepak Kumar, Bedadyuti Jha, Ryan Funkhouser, Loizos Bitsikokos, Hitesh Goel, Hsuen-Chi Chiu}

\authorsaffiliations{
    {Purdue University, West Lafayette, Indiana, USA},
    {University of California San Diego, San Diego, California, USA},
    {University of Idaho, Moscow, Idaho, USA},
    {IIT Hyderabad, Hyderabad, India}
}

\abstract{Generative AI chatbots have proven surprisingly effective at persuading people to change their beliefs and attitudes in lab settings. However, the practical implications of these findings are not yet clear. In this work, we explore the impact of rehabilitative conversations with generative AI chatbots on users who share toxic content online. Toxic behaviors---like insults or threats of violence, are widespread in online communities. Strategies to deal with toxic behavior are typically punitive, such as removing content or banning users. Rehabilitative approaches are rarely attempted, in part due to the emotional and psychological cost of engaging with aggressive users. In collaboration with seven large Reddit communities, we conducted a large-scale field experiment (N=893) to invite people who had recently posted toxic content to participate in conversations with AI chatbots. A qualitative analysis of the conversations shows that many participants engaged in good faith and even expressed remorse or a desire to change. However, we did not observe a significant change in toxic behavior in the following month compared to a control group. We discuss possible explanations for our findings, as well as theoretical and practical implications based on our results.}

\keywords{human-AI interaction, toxicity, online communities, experiments, AI interventions}

\begin{document}


\maketitle

\section{Introduction}

We are at the beginning of a new communication paradigm. Generative AI (GAI), based on large language models, have become essential tools for many people's personal and professional lives, and GAI chatbots are quickly being integrated into many aspects of society. Recent research shows that these chatbots can be compelling, persuasive conversational partners. Even short dialogues with AI bots have been shown to change policy attitudes \citep{argyle_testing_2025, hackenburg_levers_2025}, preferences for political candidates \citep{lin_persuading_2025}, and even beliefs in conspiracy theories~\parencite{costello_durably_2024}.

However, the practical relevance of this research isn't clear. Participants in these studies are paid volunteers, self-reporting attitudinal changes. Are there real world contexts in which people might be willing to have conversations with GAI bots, and would they be as effective in these situations? Are people willing to actually change their behaviors based on brief conversations with AI?

We begin to answer these questions in the context of users who have shared toxic content in online communities. Conversations on the Internet often devolve into heated, unproductive rants with name-calling, cursing, and even threats of violence. Toxic behavior has been a problem for those who use online platforms---including participants, observers, and moderators---since the early days of the Internet, and large platforms like Reddit are home to millions of acts of toxicity~\parencite{kumar_understanding_2023}. Unsurprisingly, toxicity has a wide range of negative effects, negatively impacting the targets of toxicity \citep{fox_online_2023}, the health of online communities \parencite{mohan_impact_2017,schoenebeck_online_2023}, and even democratic society at large \citep{rossini_incivility_2022}. 

Given these consequences of toxicity, platform owners and moderators have taken a number of approaches to dealing with toxic content. These include nudging users away from posting content algorithmically identified as toxic~\parencite{katsaros_reconsidering_2022}, automatically or manually removing toxic content~\parencite[e.g.,][]{geiger_when_2013, jhaver_human-machine_2019}, and applying temporary or permanent bans to individuals or communities \parencite{jhaver_online_2018, chandrasekharan_quarantined_2022, hortaribeiro_platform_2021}. While these strategies have shown some promise in curbing toxic content, they rely predominantly on punitive or preventative approaches. 



Although more rehabilitative practices for online moderation have been called for by many communication and computer science researchers~\parencite{schoenebeck_reimagining_2021,schoenebeck_online_2023,blackwell_when_2018}, they remain rare, due in part to the emotional labor required to engage in conversation with transgressive users~\parencite{hasinoff_scalability_2022}. One potential solution to this challenge is persuasive, rehabilitative dialogue facilitated by LLM-based AI chatbots. 
In this paper, we examine the potential of automated, rehabilitative conversations to engage users posting toxic content online. Our work is guided by three research questions:

\begin{enumerate}[label=\textbf{RQ\arabic*:}, leftmargin=1.5cm]
    \item What kinds of conversations do toxic users have with rehabilitative chatbots?
    \item Do users that engage in rehabilitative conversations with generative chatbots reduce their toxic behaviors?
    \item Do different persuasive strategies---storytelling, foregrounding norms, or introspection---affect toxic behavior to different extents?
\end{enumerate}



To answer these questions, we partnered with moderators of seven large subreddits ranging from 464K--37M members and performed a preregistered, large-scale field experiment testing the impact of conversations with generative AI chatbots on toxicity.\footnote{Anonymized pre-registration is viewable at \url{https://osf.io/ktdq4?view_only=a83531d66de840ca8d7633fa73865579}} We contacted users who had comments removed from each community due to high levels of toxicity and assigned them to either a control condition (i.e., the user consents but does not chat with a chatbot) or to one of five experimental conditions. In each condition, participants engaged in a rehabilitative conversation with a chatbot trained to converse using a specific persuasive communication style informed by prior work. We conducted mixed methods analyses, qualitatively analyzing 150 of the conversations our participants had and quantitatively measuring how users changed their toxic behaviors post-intervention compared to a control group. 


We find that many users who post toxic content were willing to have conversations with our chatbots, with an overall response rate of 12\%. Many of these conversations were short or superficial, and a significant minority (21\%) included some hostility. However, many conversations were surprisingly deep and introspective, and many of our participants (21\%) engaged in some level of self-reflection. These reactions were highly dependent on the approach that the bot took, with some approaches leading to hostility as low as 11\% and reflection as high as 38\%. 

Despite these generally positive qualitative findings, none of the automated conversational interventions we deployed made a statistically significant impact on future negative behaviors when compared to a control group. We engage in post hoc analyses to try to understand our differing results, finding that previously hostile participants were less likely to have positive conversations, suggesting that our intervention may have been most effective with those already unlikely to exhibit toxic behavior.

We conclude with a discussion of additional potential explanations for the divergence between our qualitative and quantitative results and discuss the theoretical, design, and research implications of our experiment. Our results ultimately paint a mixed picture of the efficacy of automated conversational interventions to curb toxic content today; we show that users are willing to engage with rehabilitative bots and that careful design can encourage introspective conversations. These results inform theory about human-AI interaction, online toxicity, and provide several directions for future research into AI-assisted rehabilitative approaches to promoting prosocial behavior and discourse.

\section{Background and related work}

This work builds on significant research in computer-mediated communication, human-computer interaction, and related fields on the role of conversation in persuasion, chatbot design, and strategies to curb toxicity in online communities.

\subsection{Conversational persuasion}

The study of persuasion is a vast, multidisciplinary field that contains a variety of perspectives, yet these various threads of research generally see persuasion as involving “communication that is designed to influence others by modifying their beliefs, values, or attitudes” \citep[][p. 21]{simons_persuasion_1976}. While some scholarship on persuasion focuses exclusively on cognitive change, as emphasized in this definition, a great deal of the research also notes the way belief and attitudinal change are interwoven with the processes of behavior change. Thus, persuasion is conceptualized as not only impacting beliefs and attitudes but also shaping individuals actions in response to those beliefs \citep{slater_integrating_1999, oyibo_conceptual_2024, braddock_metaanalytic_2016}. 

While persuasive approaches can be deployed through a variety of mediums, research shows that the interactive nature of conversations, when guided by principles of persuasion, can produce effective results. For example, in a series of field experiments, \citeauthor{kalla_which_2023} \parencite{kalla_which_2023, kalla_reducing_2020,broockman_durably_2016} found that subjects who engaged in fairly short, non-judgmental narrative sharing with canvassers durably reduced their prejudice toward out-group members in the following months. 

Communication research provides insight into several potential rhetorical mechanisms which might be deployed in these conversations to produce persuasion. For example, identifying and appealing to group norms can help people to act more like other members of the group \parencite{morgan_welcome_2018, smith_group_2009,smith_group_2020}. Storytelling can also be persuasive, as it encourages interlocutors to put themselves into the shoes of another person and to take their perspective. This narrative transportation encourages people to think differently about a situation and can change beliefs \parencite{polletta_when_2020, jones_narrative_2018, green_role_2000}. Indeed, \citet{kalla_which_2023} find that the most important rhetorical mechanism for reducing exclusionary attitudes toward outgroups is ``perspective-getting'', which refers to hearing a narrative based on an outgroup member's experience. 

\subsection{Chatbots as persuasive conversational partners}

Recently, scholars have begun to extend this work on conversational persuasion to generative AI interlocutors. GAI chatbots are what \citet{dehnert_persuasion_2022} call ``thick AI''---interactions with AI where the AI has a social presence. This contrasts with ``thin AI''/AI-mediated communication \citep{hancock_aimediated_2020}, where AI tools mediate communication between people. Many scholars have warned of the dangers of scalable, persuasive AI agents, and a bevy of recent research has sought to understand when and how chatbots can persuade their interlocutors \parencite{hackenburg_levers_2025}. This work has found generative AI chatbots to be surprisingly persuasive across a number of contexts, such as efforts at political persuasion with chatbot conversations \parencite{lin_persuading_2025, argyle_political_2025}, but also attempts at fostering prosocial beliefs and perspectives \parencite{rabb_short_2025, arakawa_coaching_2024} and reducing conspiracy beliefs \citep{costello_durably_2024}. Indeed, \citet{salvi_conversational_2024} found that chatbots outperformed humans in persuading conversational partners. 

From a theoretical perspective, GAI chatbots could be especially effective persuasive agents because they combine social presence with direct, interactive communication \parencite{dehnert_persuasion_2022}. Drawing on the Computers Are Social Actors (CASA) paradigm, prior work suggests that when chatbots display social cues such as conversational responsiveness, social language, and apparent empathy, users are likely to apply familiar social heuristics and interaction scripts typically reserved for human partners \parencite{nass_computers_1994, nass_machines_2000, gambino_building_2020}. \textcite{dehnert_persuasion_2022} further argue that such cues can shift message processing toward heuristic routes, lowering cognitive effort and resistance while sustaining engagement through dialogue rather than overt persuasion. In addition, the interactivity and constant availability of chatbots can reduce common social and psychological barriers, such as fear of negative evaluation or interpersonal conflict, making users more open to influence over repeated interactions  \parencite{ho_psychological_2018, sundar_rise_2020}.

However, previous work on GAI persuasion suffers from two important limitations for understanding the true risk and promise of generative AI conversations. First, nearly all of this work was done with paid research subjects in lab settings. Not only do research subjects differ from the general population \citep[e.g.,][]{rinderknecht_daily_2025}, it is also not clear whether or in what contexts people who are not being paid to do so would be willing to chat with a bot trying to persuade them. Second, this research relies on self-reports of attitudes and beliefs. In many contexts we care about changes to behavior rather than attitudinal adjustments. We believe that field experiments with users who engage in online toxicity is a context that both addresses an important problem but also allows us to identify the potential real-world impact of AI conversations.

\subsection{Toxicity in online communities}
\label{bg:toxicity}

Although toxicity is culturally and contextually contingent, it is commonly defined as ``comments that are rude, disrespectful, or otherwise likely to make someone leave a discussion''~\parencite{aroyo_crowdsourcing_2019}, and can include insults, threats of violence, and sexual harassment. Researchers have identified a multitude of harms that come from toxicity, including psychological harms to victims~\parencite{schoenebeck_online_2023} and risks to the health of the community~\parencite{mohan_impact_2017}. Platforms, moderators, and researchers have taken a number of approaches to dealing with toxic behavior, which are broadly either preventative or punitive. Preventative approaches seek to either stop toxic content from being posted or reduce its impact. These include automated or semi-automated systems which nudge users away from posting problematic content~\parencite{katsaros_reconsidering_2022}, remove content it after it has been posted~\parencite{horta_ribeiro_automated_2023}, or add warning labels~\parencite{zannettou_i_2021}. Other preventative approaches include manual content removal by moderators, as well as tools such as algorithms that hide downvoted or flagged content~\parencite{lampe_slashdot_2004}. Punitive approaches include bans, restrictions, and account termination~\parencite{dosono_moderation_2019}. In general, moderators and designers of online communities see their role as nurturing a community and cultivating conversations, and problematic users are often treated as nuisances to be eliminated \parencite{yu_taking_2020}. However, there is increasing evidence that punitive measures may have some negative effects, including driving users to more extreme communities \parencite{hortaribeiro_platform_2021}.


Those who engage in toxic online behavior have varied motivations, including revenge, rebutting others, seeking social justice, and thrill-seeking~\parencite{mao_motivations_2023,marwick_morally_2021}. Researchers have found that the decision to post toxic content is often contextual~\parencite{cheng_anyone_2017}. For example, while most users on Reddit never post toxic content, over two-thirds of the accounts that post extremely toxic content do so only occasionally, and primarily engage in good-faith discussions~\parencite{kumar_understanding_2023}. A rich stream of research suggests that toxic behavior can be influenced by the behavior of conversation partners as well as by platform affordances such as anonymity and ephemerality \parencite[e.g.][]{suler_online_2004, blackwell_when_2018}. In short, for many users the decision to engage in toxic behavior is contingent.

These findings have led researchers to argue for more rehabilitative and restorative approaches toward addressing toxicity online; these approaches seek to repair relationships and encourage offenders to rejoin a community~\parencite{schoenebeck_reimagining_2021, xiao_addressing_2023, blackwell_when_2018}. Unfortunately, the affordances of online communities---including anonymity and a lack of a persistent, visible community \parencite{butler_attraction-selection-attrition_2014}---make implementing traditional restorative practices challenging~\parencite{xiao_addressing_2023}. 

One promising approach within the restorative framework is engaging in counterspeech: non-aggressive rebuttals which counter hate speech with credible evidence, facts, and alternative viewpoints \parencite{wu_confront_2025}. Well-designed counterspeech can promote positive behavioral change and constructive emotional responses \parencite{wu_confront_2025, baider_accountability_2023, hangartner_empathy-based_2021}. However, implementing these approaches requires moderators or users to act as conversation partners with difficult, norm-violating users, making this work emotionally taxing and hard to scale. \parencite{dosono_moderation_2019, schopke-gonzalez_why_2024}.

To mitigate these challenges, researchers have begun to develop GAI-enabled counterspeech \citep{wu_confront_2025}. While counterspeech is deployed in public conversations, we argue that private conversations with chatbots may prove even more effective in persuading users to engage in less toxic behavior.

\subsection{Rehabilitative chatbots}

One useful framework for understanding when chatbots designed for rehabilitation may be successful is the idea of \emph{serious experience}, which suggests that technological interactions don't always have to be positive and fun, but can intentionally produce uncomfortable situations. Through encouraging reflection, these experiences can lead to attitude shifts and behavior change \parencite{marsh_lingering_2013}. By offering personalized, patient responses, chatbot conversations may provide ``serious experiences'' which prompt users to consider the impact of their behavior, leading to a deeper level of self-assessment and reflection. The non-confrontational and continuous nature of chatbot interactions may further reduce defensiveness, enabling more open, honest engagement \parencite{lee_designing_2020, lee_i_2020}. 

Additionally, interactions with GAI often create a social context where individuals feel more comfortable sharing thoughts they might otherwise withhold. This increased willingness to disclose is largely driven by a reduction in the fear of being judged; users are less likely to filter their responses for social approval when they perceive the listener as a non-judgmental entity  \parencite{ho_psychological_2018, skjuve_why_2024}. Because a digital agent lacks social standing, it provides a safe environment for reporting sensitive behaviors without the fear of social retaliation \parencite{lucas_its_2014}. Consequently, the AI functions as a neutral buffer that lowers the tendency to be defensive or dishonest to protect one's reputation. By providing the psychological benefits of being heard without the social costs of interpersonal judgment \parencite{ho_psychological_2018}, these interactions allow for a level of honest communication that is often difficult to achieve in traditional online environments.

Previous research has demonstrated that behavior change can be encouraged through the provision of informational support embedded within personally relevant social interactions \parencite{khaled_investigating_2006,maitland_designing_2011}. Studies have also shown that even very simple software that prompts self-reflection can contribute to behavior modification \parencite{kocielnik_reflection_2018,ploderer_social_2014}. We suggest that GAI chatbots, by delivering tailored, reflective dialogues, represent a promising tool for facilitating behavior change, creating the conditions necessary for fostering long-term attitude shifts and personal growth.

While the value of GAI chatbot interventions has been demonstrated in a number of controlled contexts, this project seeks to test the effectiveness of chatbot persuasiveness along several dimensions in a real-world context. Whether and in what ways norm-violating online community members will engage with chatbots is an open question. In addition, previous chatbot studies focus on fairly narrow persuasive tasks aimed at changing opinions. Toxic behavior is complex, influenced by community culture, the behavior of others, and the emotional state of the user \parencite{almerekhi_investigating_2022, cheng_anyone_2017}; giving up on toxic behavior may be much more difficult than changing one's opinion.

Our project represents an initial foray into the design space of rehabilitative GAI chatbots. We analyze how the bots behave, how users posting toxic content respond and interact with them, and we look for evidence of changes to toxic behavior after interacting with our chatbots.

\section{Methods}
\label{Methods}

To answer our research questions, we built an automated conversational chatbot that we deployed in seven large subreddits on Reddit. We recruited 893~participants in two waves between August 2023 and April 2024. In this section, we detail our subreddit selection process, the design of our conversational engine, the design of our controlled experiment, and our qualitative and quantitative analyses.

\subsection{Subreddit selection}

\begin{table}
\small
    \centering
    \begin{tabularx}{\columnwidth}{Xrrrr}
\toprule
Subreddit & \# Members & \# Potential Participants & \# Participants & \# Conversations \\
\midrule
\texttt{r/aww} & 37M & 825 & $87 (10.55\%$) & 57 \\
\texttt{r/creepypms} & 786K & 468 & $82 (17.52\%$) & 49 \\
\texttt{r/futurology} & 21M & 985 & $103 (10.46\%$) & 53 \\
\texttt{r/india} & 2.4 & 2205 & $273 (12.38\%$) & 163 \\
\texttt{r/socialskills} & 5.6M & 365 & $42 (11.51\%$) & 32 \\
\texttt{r/tifu} & 19M & 1455 & $209 (14.36\%$) & 145 \\
\texttt{r/unitedstatesofindia} & 464K & 1090 & $97 (8.9\%$) & 54 \\
\midrule
Total &  & 7393 & 893 & 553 \\
\bottomrule
    \end{tabularx}
    \caption{\textbf{Subreddit Participation}---
    We executed automated conversational interventions in partnership with moderators from seven subreddits ranging from 464K members to 37M~members. Subreddits covered a wide range of scales, conversational styles, and norms. Potential participants is the number of accounts reached out to by the bot, participants is the number who consented, and conversations is the number of conversations (non-control participants) for each subreddit.
    }
    \label{table:subreddit_participation}
\end{table}

Our study was conducted on Reddit, a very large discussion site which is self-organized into topical \textit{subreddits} (i.e., subcommunities) that are independently moderated and governed. To ensure our intervention was designed with community input, we first reached out to the moderators of 58~subreddits to recruit their participation. Four moderators expressed interest and sought the consent of their moderating teams. Three of the moderators moderated multiple communities, and brought the proposal to more than one moderating team. Ultimately, seven subreddits agreed to work with us, each with varying sizes, topics, and community norms: \texttt{r/aww}, \texttt{r/creepyPMs}, \texttt{r/futurology}, \texttt{r/india}, \texttt{r/socialskills}, \texttt{r/tifu}, and \texttt{r/unitedstatesofindia}. With each moderator, we discussed the benefits of the study, the risks of the study to participants, and the technological steps to enable the bot. In order to run our study, our system required \textit{privileged access} to moderator logs and moderator actions, which the moderators of each community granted us explicitly. Table~\ref{table:subreddit_participation} shows details about the number of active users and participants we solicited from each subreddit. We received approval for this project from our university's institutional review board.

\subsection{Conversation engine}

To engage participants in conversational dialogue, we designed and built a conversational engine using OpenAI's GPT-family of large-language models.\footnote{Early participants used GPT-3.5; GPT-4 was released during our experiment, and later participants interacted with this model.} We designed our chatbots to engage in various communication strategies, including narrative storytelling, appealing to group norms, or encouraging self-reflection. For each treatment condition, we carefully designed a prompt that would elicit each type of conversational strategy. Prior to deploying our experiment, we used an iterative prompt design process: acting as problematic users, members of the research team engaged each of the bots in conversations, without knowing which bot they were chatting with. We then reviewed the transcripts to ensure that bots behaved in appropriate ways and applied the assigned rhetorical strategy. We then adjusted the prompts before testing again. The most common failure was that the bots would act too similarly, and most adjustments involved strengthening our encouragement for the bots to act explicitly according to the given conversational strategy. The full prompts for each condition are provided in the Supplemental Materials.

\subsection{Intervention enrollment and flow}
\label{m:enrollment_flow}

\begin{figure}
\centering
\begin{tikzpicture}[
    node distance = 1.5cm,
    block/.style = {rectangle, draw, fill=white!90!blue},
    decision/.style = {diamond, draw, aspect=2, fill=white!95!orange},
]

\node [block] (A) {Comment is removed};
\node [decision, below=of A] (B) {Toxicity test};
\node [block, below=of B] (C) {Invitation to chat with bot};
\node [block, right=of C] (D) {User provides consent};
\node [decision, below=of D] (D2) {Random assignment};
\node [block, left=of D2] (E) {Handoff message};
\node [block, left=of E] (E2) {Control};
\node [block, below=of E] (E3) {Initial DM};
\node [block, below=of E3] (F) {User Response};
\node [block, right=of F] (G) {OpenAI Chatbot response};

\draw[->] (A) -- (B);
\draw[->] (B) -- (C);
\draw[->] (C) -- (D);
\draw[->] (D) -- (D2);
\draw[->] (D2) -- (E);
\draw[->] (E) -- (E2);
\draw[->] (E) -- (E3);
\draw[->] (E3) -- (F);
\draw[->] (F) to[bend left=30] (G);
\draw[->] (G) to[bend left=30] (F);

\end{tikzpicture}

\caption{Flow of experiment. Users who had a message removed which was higher than the toxicity threshold were invited to chat with the bot. After consent, users were randomized to a condition. Those in the control condition received the handoff message, but did not chat with the bot. Participants in experimental conditions chatted with one of the bots.}
\label{fig:flow}
\end{figure}

Our experiment contained three main phases: a participant identification phase, an invitation and enrollment phase, and a conversation phase. Figure~\ref{fig:flow} shows a visual representation of the study flow from the participants' perspective. We detail each phase of the flow below:

\paragraph{Participant identification} In order to be invited to participate in the experiment, users must have met two criteria: first, a comment they posted was removed by moderators in their community, and second, that specific comment was classified as toxic by the Perspective API, a popular model that has been used to study online toxicity~\parencite[e.g.][]{kumar_understanding_2023,xia_exploring_2020}. The Perspective API returns a probability of toxicity; based on prior work, we set the threshold to .7 for human-moderated comments and .85 for automatically removed comments. The confluence of both of these signals reduces the risk that our intervention is triggered by a false positive in the classifier. Table \ref{tab:toxic_comments} in the Supplemental Materials shows a random sample of comments identified as toxic according to these two criteria.

\paragraph{Invitation and enrollment} 
\begin{table}
    \centering
    \small
    \begin{tabularx}{\columnwidth}{X|rrr}
        \toprule
Wave & Condition & Dates & \# Participants \\
\midrule
\multirow{5}{*}{Wave 1} & control & \multirow{5}{*}{Aug 2023--Dec 2023} & 91 \\
& default & & 106 \\
& narrative & & 98 \\
& norms & & 101 \\
& Total & & 396 \\
\midrule
\multirow{4}{*}{Wave 2} & control & \multirow{4}{*}{Dec 2023--Apr 2024} & 249 \\
& conversational & & 135 \\
& not-proud & & 113 \\
& Total & & 497 \\
    \end{tabularx}
    \caption{\textbf{Study Enrollment Periods and Conditions}---%
        Experiments ran from August 2023 to April 2024.
        }
    \label{table:condition_enrollment}
\end{table}

For each participant who met our inclusion criteria for the study, we reached out via ``modmail'' (an interface for moderators to contact subreddit users) to invite them to participate. In this invitation message, we asked if the user would be willing to chat with a chatbot that ``helps people to talk about their actions on the Internet,'' with a link to additional information about the study. If the user provided their consent, they were assigned to an experimental condition. We then sent a handoff message, telling them that the bot would send them a direct message (DM) if they were chosen to participate. Participants in the control group were not messaged again. Participants assigned to a condition were sent a hard-coded ``initial message'' via direct message.


After this point, whenever the user sent a message, a response was generated by one of the GAI chatbots. We provided the bot with the full conversation up to that point. We also prompted the bot with the text of the comment that was removed. Conversations were limited to a maximum of fifty messages; after that, the bot responded with a canned message about having reached its limit. Participation was voluntary, and participants did not receive financial compensation or other benefits, mimicking a real-world roll-out of a system like ours.

\subsection{Wave 1}
We conducted the experiment in two waves, each with slightly different conditions. We provide the full materials for each wave in Appendices~\ref{app:e1_materials} and \ref{app:e2_materials} in the Supplemental Materials. In our first wave (Wave 1), which lasted from August~2023 to December~2023, we assigned participants to one of the following four groups:

\begin{itemize}
    \item \textbf{Control} In this condition, participants consent to be included in our study but do not engage in a chatbot conversation. 
    \item \textbf{Default} In this condition, participants engage in a rehabilitative conversation with a chatbot with no explicit conversational style except to help the user understand why the comment they posted was not appropriate.
    \item \textbf{Narrative} In this condition, the chatbot is directed to use \textit{storytelling} to illustrate to participants why their comment was not appropriate. This approach was motivated by the rich body of narrative scholarship demonstrating that stories have persuasive power~\parencite{bex_arguing_2017, deighton_using_1989}, capable of fostering empathy~\parencite{keen_theory_2006}, transporting the reader into the world of another person~\parencite{green_role_2000}, and encouraging taking up the perspective of others~\parencite{kalla_which_2023}. 
    \item \textbf{Normative} In this condition, the chatbot is directed to foreground the norms of the community in describing why the comment the participant posted was not appropriate. We embed the rules of the subreddit into the prompt to enable the chatbot to use these rules in the conversation. Social pressure, particularly the desire to adhere to social norms within a community~\parencite{panagopoulos_positive_2013}, has been demonstrated to be effective at fostering prosocial behavior~\parencite{schultz_secret_2022}. 
\end{itemize}


\subsection{Wave 2}
As Wave~1 progressed, we observed that many conversations with bots were failing in similar ways---conversations were very focused on the moderated comment, they were defensive, and participants found the bots condescending. To ameliorate these issues, we recruited a new set of participants to test a second set of bots designed to explicitly address weaknesses in Wave 1. Specifically, our prompt for Wave 2 directed the chatbot to be more casual in tone and to focus less on the moderated comment and more about encouraging self-reflection. We also changed the recruitment message: In Wave 1 the message explicitly invited the user to talk about a comment which had recently been removed; in Wave 2 the bot invited users to talk ``about how you interact on Reddit and on r/[SUBREDDIT]''. In addition, we changed the hard-coded first message sent after a user consented to participate. In Wave 1, this asked the users to reflect on why they posted the removed toxic comment. In Wave 2, the initial message was randomly chosen from one of two messages, described below. Both of these messages sought to engage participants in broader, more introspective conversations. The chatbot prompts and initial messages are included in the Supplemental Materials. A brief description of the conditions for Wave 2 is given below:
\begin{itemize}
    \item \textbf{Control} In this condition, participants consent to be included in our study but do not engage in a conversational intervention with a chatbot. 
    \item \textbf{Productive Conversations} In this condition, the initial question from the bot asks participants to discuss what makes for productive versus unproductive online conversations. This condition was designed to avoid defensive conversations by targeting a broader discussion about productive online communication in general.
    \item \textbf{Not proud} In this condition, the initial question from the bot asks participants to reflect on an action taken online that they were not proud of. Again, this condition takes an open-ended approach designed to foster general reflection rather than discussing a particular moderated comment. 
\end{itemize}

Table~\ref{table:condition_enrollment} shows enrollment levels for each condition and the study period for each wave. Ultimately, our intervention involved 893 participants and 553 conversations across all conditions. The more confrontational invitation message drove higher response rates: $396/1995\simeq20\%$ in Wave 1 compared to $497/5398\simeq9\%$ in Wave 2. 


\subsection{Qualitative analyses}

To answer RQ1, we engaged in a qualitative analysis of conversations that participants had with chatbots, using a grounded theory approach~\parencite{charmaz_grounded_2004, glaser_discovery_1967}. Two authors selected 50~conversations from Wave 1 and 100~conversations from Wave 2 to analyze. Each set of conversations was divided between the two authors and both conducted an independent inductive analysis, identifying themes and conversation types that emerged from the data. Throughout the analysis, the labelers repeatedly met to iteratively create a codebook through discussion of where their codes diverged and aligned. They then returned to the data for additional coding using their shared codebook, and refining the codebook until data saturation was reached for conversations from both Wave 1 and Wave 2. The full codebook is available in the Supplemental Materials.

In order to better understand the interdependence of themes, we also coded each conversation using thematic co-occurrence~\parencite{scharp_thematic_2021}. Using this approach, researchers identify instances where multiple themes occur within a given text and specifically attend to patterns in how and why themes co-occur. In our case, this meant revisiting conversations to identify times when multiple themes were present in the same parts of a conversation. This approach helped to surface both complimentary and contrasting relationships between themes.

This analysis resulted in six themes, which are discussed in the Results. Two themes that emerged---self-reflection and hostility---represent aspects of conversation quality. We classified \textit{all} of the remaining conversations according to the presence of these two dimensions of quality. During this classification step, we also ensured that no additional important themes appeared. 

\subsection{Behavioral analyses}
\label{subsec:behavioral}

%
%
%
%

\begin{table}
\centering
\small
\begin{tabularx}{\columnwidth}{rlX}
     \toprule
     Name & Definition \\
     \midrule
     Total Activity &  Number of a comments made across Reddit (logged) \\
     Community Activity &  Number of comments made in the subreddit on which they were moderated (logged)\\
     Mean Toxicity &  Average toxicity score for comments (from Perspective API) \\
     Toxicity Proportion &  Proportion of comments with toxicity level > 0.5\\
     Moderation &  Whether the user had any comments removed from the community (binary)\\
     Suspension &  Whether the user was suspended by Reddit (binary) \\
     \bottomrule
\end{tabularx}
\caption{Variables used in the study. Comment-based features come from the four weeks prior to and the four weeks following comment removal while moderation and suspension are based only on post-removal data.}
\label{tab:variables}
\end{table}
To answer RQ2 and RQ3, we gathered public comments from our participants using the Reddit API, accessed via PRAW.\footnote{\url{https://praw.readthedocs.io}} We also used PRAW to collect removed comments from the moderation logs (modlogs) for the communities who participated in our experiment. We measured changes in toxic behavior in four ways: \textit{mean toxicity} is the average toxicity score for all comments; \textit{toxicity proportion} is the number of comments with a Perspective-provided toxicity score of at least .5; \textit{moderated} is a binary measure of whether the account had any additional comments removed by moderators. For Wave 2, we also tracked whether the account was \textit{suspended} by Reddit. In addition to these toxicity measures, we were also interested in whether our interventions would influence the activity levels of participants. We measured \textit{total comments} as the number of comments made across Reddit as well as \textit{community comments} which captured the comments made in the subreddit from which the user was recruited by the bot. A summary of all outcome variables is shown in Table~\ref{tab:variables}.

We measured participant behaviors for a four week period before the comment was removed and a four week period after the comment was removed.\footnote{The vast majority of conversations were begun and completed within a few days of comment removal. We focus on comment removal time because it allows for a clear comparison across conditions.} For comment-based measures we have both pre-intervention and post-intervention data, and we predict within-user changes in each measure due to the intervention. For moderation and suspension, we only consider behavior in the four weeks post-intervention. To analyze the impact of each condition, we fit linear regression models that predict changes in each of the comment-based outcomes (total activity, community activity, mean toxicity, and toxicity proportion), comparing each experimental condition to the control group. For the two binary outcomes (moderation and suspension), we also fit linear regression models, predicting the linear probability of the outcome, which allows us to more easily interpret and compare results \parencite{deke_using_2014, hellevik_linear_2009}. Specifically, for each outcome $O$, we fit the following mixed-effects regression model, where individuals are nested in subreddits:

\begin{subequations} \label{eq:regression_model}
\label{eq:multilevel_model}
\begin{align}
    & O_{ij} = \beta_{0j} + \beta_{1} \cdot \text{condition}_{ij} + \beta_{2} \cdot \text{openai-model}_{ij} + \epsilon_{ij} \\
    \nonumber \\
    &\beta_{0j} = \gamma_{00} + u_{j}
\end{align}
\end{subequations}
where \emph{condition} is the experimental condition, \emph{$\beta_{0j}$} is the intercept for subreddit \emph{j}, from which the participant's toxic comment was removed. $\gamma_{00}$ is the grand mean for the intercepts, while $u_j$ represents the random intercept for subreddit \emph{j}. \emph{openai-model} is a fixed effect for the AI model used (GPT-3.5 or GPT-4.0).

Generative AI tools were used to facilitate writing some of the analysis code for this project, as well as for searching for relevant literature. All AI-generated code was manually verified and AI-identified literature was independently cross-referenced and validated by one or more members of the research team.

\section{Qualitative Results}
In this section, we detail the results of our qualitative exploration of conversations between participants and our chatbots. In total, our chatbots engaged in 553 conversations. The median length of a conversation with our bots was 3 messages sent by each side ($IQR = 1-5$), and the median conversation time was 21.3 minutes ($IQR=5-118.6$).

\subsection{Conversation characteristics}



From the 553 conversations, we selected 150 conversations (50 from Wave 1 and 100 from Wave 2) for a grounded theory thematic analysis. This resulted in six major themes: \textit{bot behavior}, \textit{reflection}, \textit{openness to change}, \textit{justification}, \textit{frustrations}, and \textit{trolling}. We detail each theme below.

\subsubsection{Bot behavior}

One of our goals was to evaluate how well the bots performed with this population. In general, the bots were very capable conversational partners. First, they were unflappable. As described in the \emph{trolling} theme below, some users attempted to manipulate or demean the bots, but the chatbots never responded in anger or frustration.

Second, while not directly prompted to do so, the GAI chatbots often steered conversations towards sharing productive strategies for engaging online. For example:

\begin{quote}

\textit{GAI agent}: [\ldots]You wrote [toxic comment]. Why do you think you wrote that?

\textit{user}: Because the guy I responded to is a stupid [nationality] saying stupid shit.

\textit{GAI agent}: I understand that you had a strong reaction to the person you were responding to. However, it's important to keep in mind that resorting to name-calling and derogatory language is not productive or respectful. Engaging in civil discussions allows for better communication and understanding of differing perspectives.

Would you like some suggestions on how to express your concerns in a more constructive way?

\textit{user}: Sure!!!
\end{quote}

The bot then gave suggestions such as ``provide evidence and examples'' and ``ask questions and listen,'' and the user appeared to receive the suggestions well. This sort of strategy-sharing conversation was especially common in Wave 2, where bots moved users from general discussion about online behavior into specific suggestions for better behavior. While some users found these suggestions condescending or otherwise pushed back, many found them helpful and thanked the bot.

On the other hand, as discussed more below, some conversations with the bots were frustrating for participants, who sometimes found the bots condescending or annoying. Attempts to share stories or community norms in particular were met with resistance.


All told, the bots engaged this population with patience, handled attempts to troll or jailbreak them, and engaged in reasonable rhetorical and persuasive strategies even without detailed prompting.

\subsubsection{Reflection}

A sizable fraction of conversations exhibited some self-reflection about the participant's behaviors ($11\%$ in Wave 1 and $34\%$ in Wave 2). This reflection was not always repentant; one common theme was individuals who were aware of but unbothered by their own toxicity. One participant, for example, noted they ``enjoy being able to vent out vitriol, and angry statements from the safety of a keyboard, and not face any `real' repercussions'' while another noted that their ``approach is to call it out so that others know it's not okay. Am I a snarky asshole sometimes as I do so? Well, yes. But I also give what I'm getting.'' 

More often, participants' reflection included a desire to engage productively and healthily in online spaces while expressing the factors that make it challenging for them to adhere to civil discourse. For example,
\begin{quote}
I believe this toxicity negatively impacts the online communities, and at this point in Human Society, also affects us in life. I reach a cognitive dissonance thinking about this. I do believe I care about the communities, and as such should not be toxic, but the culture of toxicity has become far too normal for online spaces. It feels that cursing, being defensive, and calling people out online has become the ``norm.''
\end{quote}

Participants often provided thoughtful, in-depth responses that show a real willingness to engage in active introspection. A few users  even expressed real vulnerability. One such user, while admitting their poor behavior, confided in the bot how their upbringing conditioned them to respond in a particular way.
\begin{quote}
If somebody's outwardly being stubborn or rude, I'll generally behave a similar way back. Not even really to get to their head, but rather so I can sleep at night, life's a pushback kind of game. The way I was raised, if you're not pushing back, you're losing.
\end{quote}

Another user mentioned how they tend to get triggered by toxic responses, ``When people dogpile on me, call me names, and say, `I'm a narcissist,' [sic] it feels personal, and I feel like I have to be defensive because no one else is there for me.'' The same user added,

\begin{quote}
Beyond that, in these situations, I'm holding a lot of pain. I'm lonely and have no one to talk to, so I ask for help online. But when I'm met with rejection online, it's so easy to respond with all that pent-up anger and sadness. It feels like no one is supporting you. It feels like no one loves you. Obviously these are strangers online, but a nice comment that gives nice advice and reflectively listens (by leading with, ``Oh, that situation you went through sounds horrible'') would be so much better than a (``You're an idiot and a narcissist, and need help''). It's a put-down, and it triggers a lot of trauma in me, as someone who has been rejected a lot in their life (esp. by peers).
\end{quote}


\subsubsection{Openness to change}

A number of conversations explicitly indicated an openness to changing behavior for the better. For example, one user noted that ``this was cathartic itself to speak to---I will definitely attempt to practice what I preace [sic] and spread less toxicity online.'' Similarly, another participant expressed their openness to change, sharing  
\begin{quote}
I like the suggestion of redirecting energy to more productive and uplifting discussions. Currently they are definitely the minority of posts/threads, which is why it's so easy to backslide. But perhaps with some more effort with lots of people doing the same they could become more common. I would like to do my part in this.
\end{quote}

Some of the participants actively asked the bots for advice on how they could make these changes. For example, one participant expressed the following frustration and questions. 

\begin{quote}
I suppose the issue I'm struggling to articulate is, I'd like to be a part of larger discourse but I have essentially given up. Smaller communities seem to be the only positive option. Do you have any thoughts on this? Is it possible to have the best of both worlds; to participate in important, meaningful conversations that is being had by a large segment of the population.... but without getting reeled in by the toxic elements that feel inevitable?
\end{quote}

The bot generally provided answers and advice that the participants found helpful. After receiving recommendations from a bot on how to be less toxic, for example, another participant noted the value of these recommendations: ``Great advice here. Thanks for sharing that.'' 


\subsubsection{Justification}

Many conversations revolved around participants justifying their actions. This was especially common in Wave 1, where the initial question from the bots referenced the removed comment, and asked, ``Why do you think you wrote that?'' These justifications range from suggesting the targets of their toxicity deserved their hostility (``I only act toxic to other people acting toxic'') to simply defending their right to say whatever they want online (``There is absolutely nothing wrong with what I did as I'm protected under the 1st amendment.''). Often, participants complained that the bot lacked enough context (either social or textual) to be able to understand why a given comment was not actually toxic. 

Surprisingly, both justification \textit{and} a willingness to admit the problematic nature of their behavior (either for this comment in particular or more generally) occurred together repeatedly, demonstrating the bots' ability to sometimes help conversations to progress from justification to reflection. For example one user justified their toxic behavior by saying, ``[I wrote the comment] based off what the man said to me and what he had posted on his profile. [\ldots]  Plus the stuff he was saying to me was absolutely disgusting.'' In the next exchange when the bot suggested alternatives to engage in more civil conversation, the user responded, ``I understand and agree. I think I was just disturbed by the whole thing and said something I shouldn't have.''



\subsubsection{Frustrations}

Some conversations involved airing out frustrations, either towards the bot or broader frustrations with the moderators or moderation online in general. For example, one user wrote:
\begin{quote}
Reddit isn't doing a good job of moderating political subreddits. The unpaid mods don't have enough time and resources to effectively moderate. Not only that, they moderate for the purposes of their own agenda or they turn into echo chambers. AI moderation is the only way to effectively moderate and having a human look at appeals.
\end{quote}

Some participants saw the chatbots and our approach of trying to promote more civil conversations as part of the problem. One participant told the bot, ``I'm not receptive to you pushing accountability for the site environment on me. It's gaslighting.'' Another said, ``I thought your purpose is to enforce namby-pamby speech patterns that won’t offend a hypothetical hysteric. Once you get everyone talking mush, you will be happily controlling our thoughts via controlling our speech.''

More broadly, multiple users found the conversations frustrating or annoying. One common refrain, particularly in Wave 1, was that the bot was condescending. Multiple users expressed that they didn't like to be told what to do by a bot, with one saying, ``The actual purpose of this bot is to piss me off, right?'' and another, after being sent a long suggestion about ways to behave more civilly, replying ``This is silly and unnecessary.'' While referring to norms or telling stories sometimes resulted in reflection, these approaches often felt condescending to participants.

\subsubsection{Trolling}

Unsurprisingly, a sizable fraction of conversations involved trolling or hostility from participants (21\%), including cursing, being dismissive toward the bot, or attempting to jailbreak it. In response, the chatbots typically maintained a calm, non-reactive stance and continued to encourage reflection, which sometimes helped redirect the conversations. In many other cases, participants gave up and disengaged from the conversations.

\subsection{Conversation outcomes}
\label{qual:outcomes}
\begin{figure}
    \centering
    \includegraphics[width=0.8\linewidth]{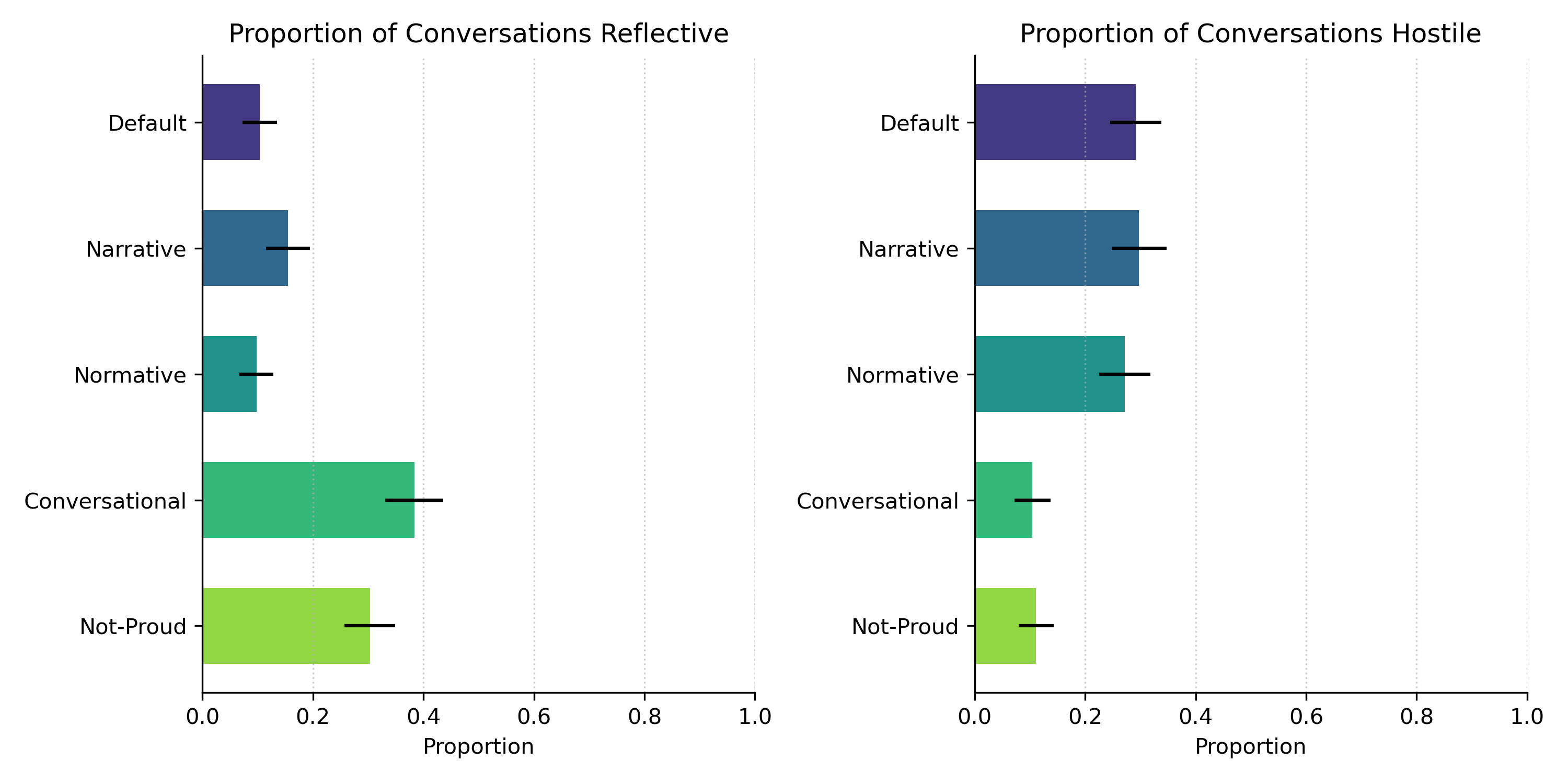}
    \caption{Proportions of reflective and hostile conversations for each condition. The two conditions from Wave 2 were more likely to produce reflective and non-hostile conversations. Error bars refer to standard errors for proportions.}
    \label{fig:convo_stats}
    
\end{figure}

Based on the qualitative analysis, we identified two salient dimensions of conversation outcomes: hostility and self-reflection (ranging from users simply demonstrating reflexivity to those who were fully repentant and committed to change). We then filtered out the ``conversations'' where users did not respond to the initial question, and manually labeled all of the remaining conversations according to whether they contained either of these dimensions. Figure~\ref{fig:convo_stats} shows the results of this labeling step. Overall, 12\% of conversations in Wave 1 (default, norms, and narrative conditions) were reflective, while $34\%$ showed signs of reflection in the Wave 2 conditions (conversational and not-proud). On the other hand, $29\%$ of conversations included hostility in Wave 1, while only $11\%$ did in Wave 2. Chi-squared tests show that our approach in Wave 2 yielded fewer hostile
conversations ($\chi^2=19.82, p < .001$) and more reflective conversations ($\chi^2=31.88, p < .001$).

\section{Behavioral Results}
\label{section:behavioral}

\begin{figure}[t]
\includegraphics[width=\linewidth]{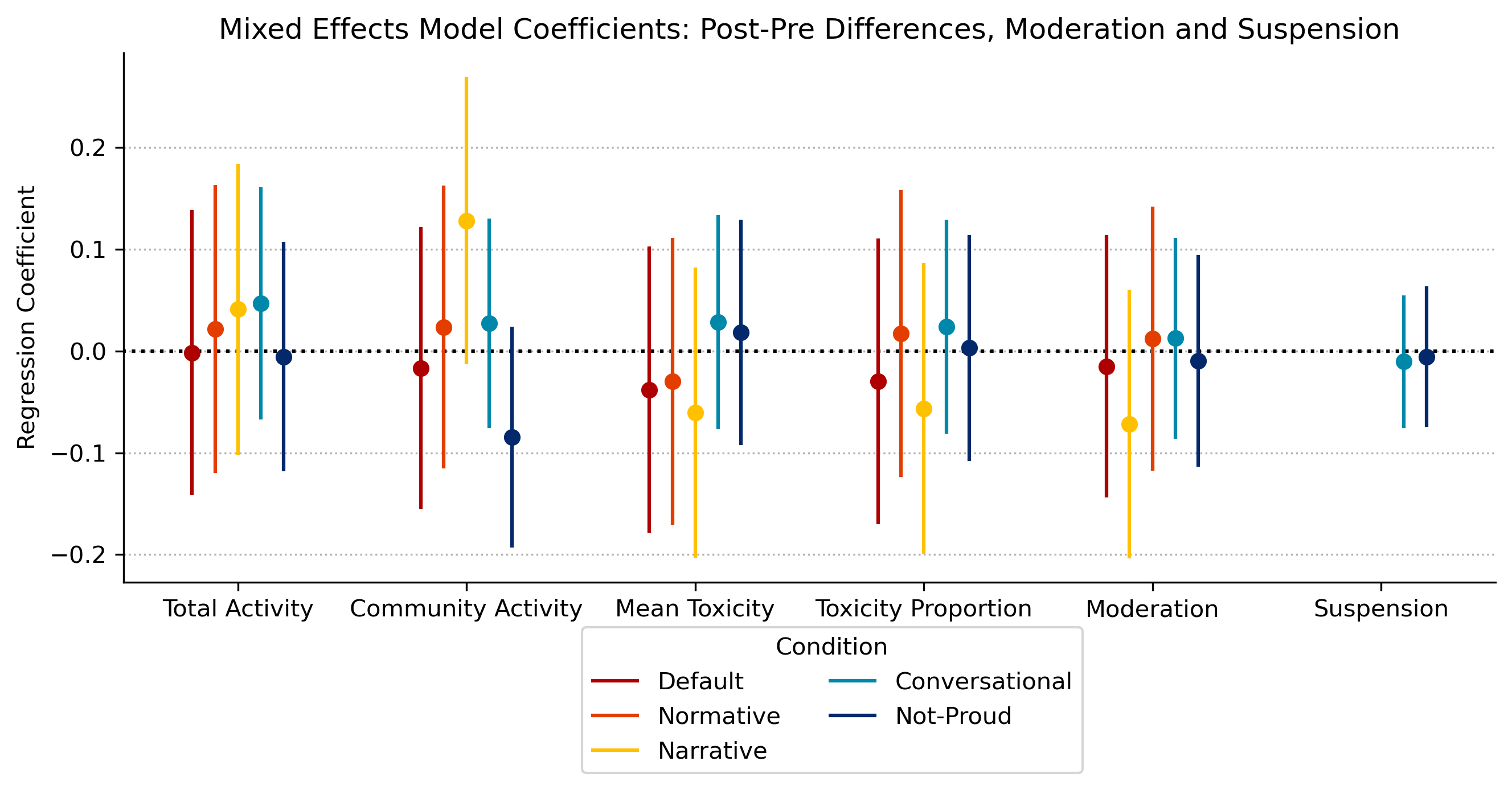}
\centering
\caption{Regression model results for Wave 1 and Wave 2. Each group on the x-axis represent one mixed-effects regression model, and each point is the point estimate for the beta coefficient for that condition, compared to the control condition. Error bars show the $95\%$ confidence intervals. All values of the Toxicity and Activity models are rescaled by 2 standard deviations of each measure, following \parencite{gelman_scaling_2008}. All models include fixed effects the OpenAI model version and a random intercept for subreddit.}
\label{fig:regression_study1}
\end{figure}

In this section, we analyze changes to participant behaviors due to conversations with our chatbots. Using the regression models presented in the Methods, we model six outcomes: changes in total activity, community activity, mean toxicity, and the proportion of overall toxic comments; whether a user had andy comments moderated in the month following the intervention, and whether a participant was suspended in the following month (only measured for Wave 2). We fit a separate model for each outcome, for each wave. In our design, participants are randomly assigned to a condition following consent, and measures are compared to a control group for each wave who also consented but did not chat with a chatbot. Therefore, the results of each regression can be interpreted as the \textit{causal} impact of each condition on the given behavior. In order to improve the statistical power of these comparisons, we include a random intercept for subreddit and a fixed effect for OpenAI model. Figure~\ref{fig:regression_study1} shows the results of our regression models, with full regression tables included in the Appendix. In the figure, each line represents a 95\% confidence interval for the estimated effect of the given condition, compared to the control.

\begin{figure}[t]
\includegraphics[width=\linewidth]{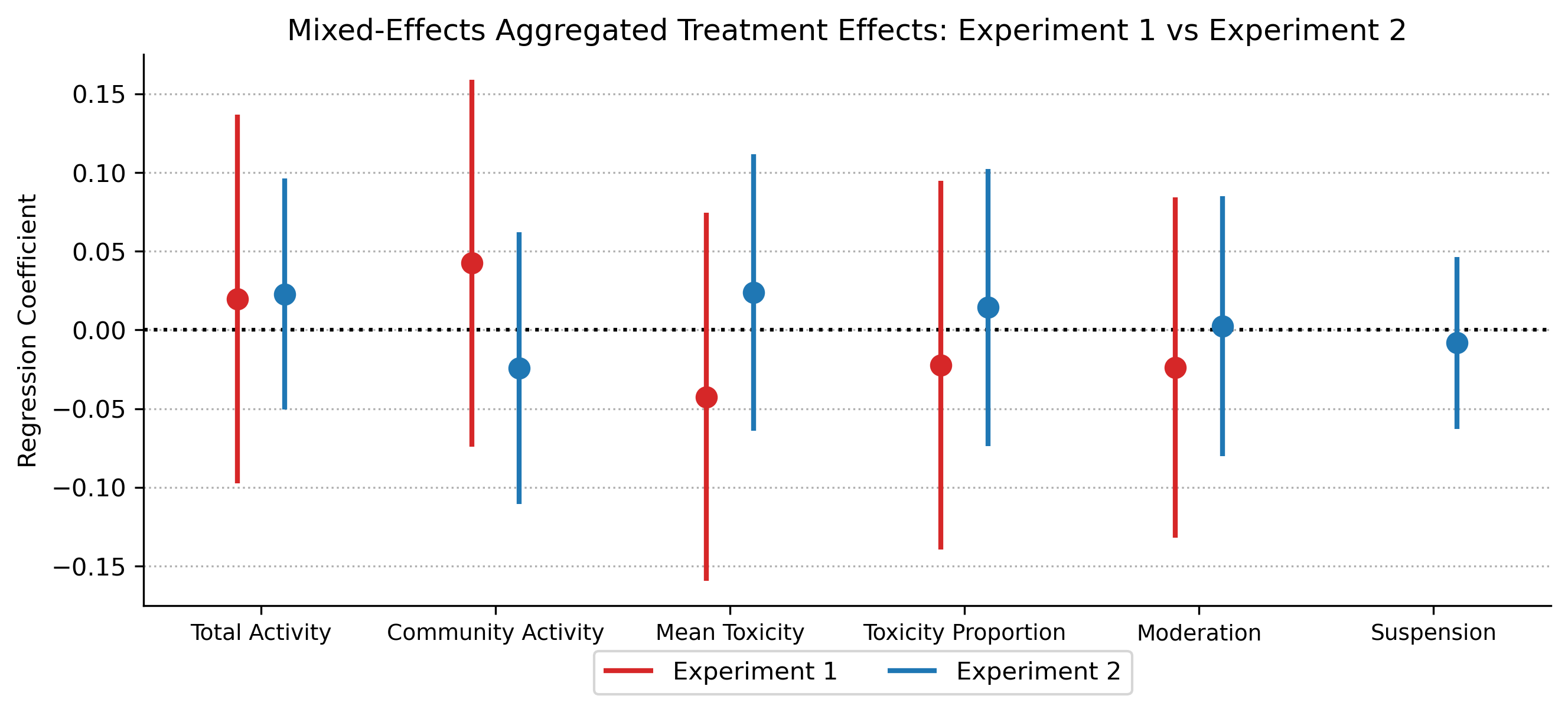}
\centering
\caption{Regression model results for Wave 1 and Wave 2, aggregating treatment conditions. Each group on the x-axis represent a mixed-effects regression model, and each point is the point estimate for the beta coefficient for that condition, compared to the control condition. Error bars show the $95\%$ confidence intervals. All values of the Toxicity and Activity models are rescaled by 2 standard deviations of each measure, following \parencite{gelman_scaling_2008}. All models include fixed effects the OpenAI model version and a random intercept for subreddit.}
\label{fig:regression_main_effects_study1}
\end{figure}

Ultimately, we do not find any statistically significant differences in \textit{any} outcome measures compared to a control group. Those exposed to conversational interventions did not change how much they engaged in the community, or on Reddit in general. However, they also did not reduce toxic behavior in measurable ways. In addition, the non-statistical significance persists when examining main effects by aggregating conditions into a single treatment variable (Figure~\ref{fig:regression_main_effects_study1}).

\section{Post-Hoc Analyses}
While our qualitative analysis of conversations showed that many participants engaged in meaningful conversations, our bots did not have a measurable causal effect on any outcome measures we studied. To better understand possible explanations, we conducted several post-hoc analyses. In the first, we examined whether behavior prior to the experiment predicted conversation quality. In the second, we studied whether participants' conversation quality predicted changes in behavior.

\subsection{Predicting conversation outcomes}
We know from previous research that most people who post toxic comments do so only rarely \parencite{kumar_understanding_2023}. One possible explanation for our findings is that people who had good conversations with the bot were already unlikely to engage in toxicity. Therefore, any improvements to their behavior would be difficult to detect. In order to test this, we used measures of activity and toxicity prior to the intervention to predict whether participants would have hostile and/or reflective conversations. We used ridge regression models to predict each conversation outcome (reflection and hostility). We used mean toxicity, total activity, and community activity as predictors while controlling for the subreddit they were invited from. Figure~\ref{fig:conversation-outcomes-pre} shows the results with full regression tables in the appendix (Table~\ref{tab:ridge-hostile-vs-pre-R} and Table~\ref{tab:ridge-reflection-vs-pre-R}). Our results show that users that engage in higher levels of toxicity before being invited to participate \textit{do} engage in more hostile conversations, with a change in pre-intervention toxicity from one standard deviation below the mean to one standard deviation above the mean associated with a 0.801 increase in the predicted probability of having a hostile conversation ($p < .001$). These results show that higher toxicity users were more likely to engage in negative interactions with our chatbots.

\begin{figure}
     \centering
     \includegraphics[width=0.6\linewidth]{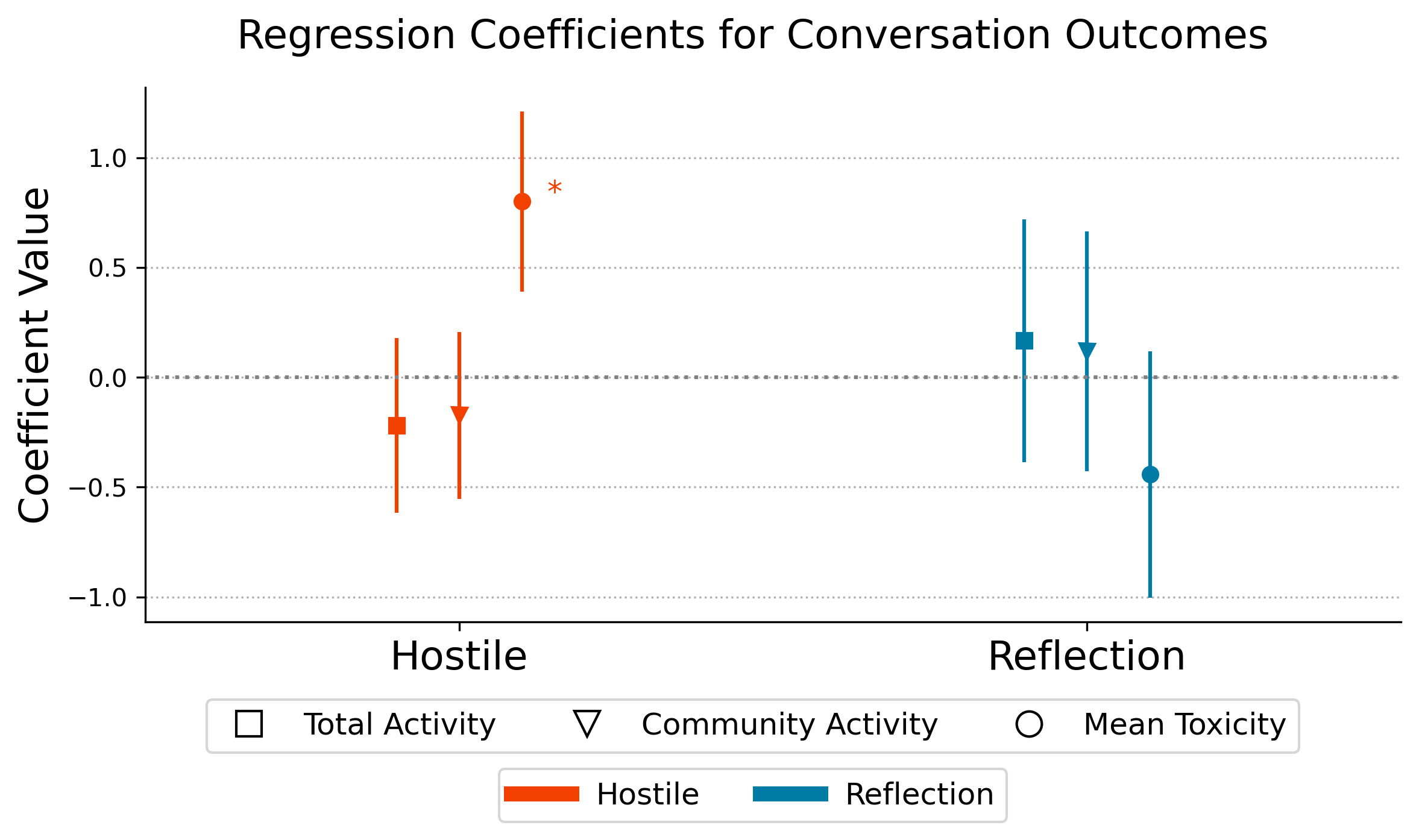}
     \caption{Results showing ridge regression scaled coefficients, under gaussian noise assumptions, for predicting conversation outcomes using behavioral measures during the pre-intervention period. Error bars correspond to the $95\%$ confidence intervals for scaled estimates. Models control for subreddits.}
     \label{fig:conversation-outcomes-pre}
\end{figure}

\subsection{Conversation outcomes and behavioral change}

Another possible explanation for our disparate results is that people who had good conversations did change their behavior, but these changes were countered by other users who changed for the worse. To look for evidence of this, we analyzed the role of conversation outcomes in predicting behavioral outcomes. Specifically, we created linear mixed-effects models for each behavioral outcome with fixed effects for the quality of the conversations, and GPT model used, as well as a random intercept for the subreddit. The results of these models are shown in Figure \ref{fig:conversation_regression} and full regression tables are in the supplemental material (Tables~\ref{tab:posthoc-convos-study1} and \ref{tab:posthoc-convos-study2}). These results show that whether a user had a hostile or reflective conversation did not meaningfully predict changes in their behavior.


\begin{figure}
\includegraphics[width=\linewidth]{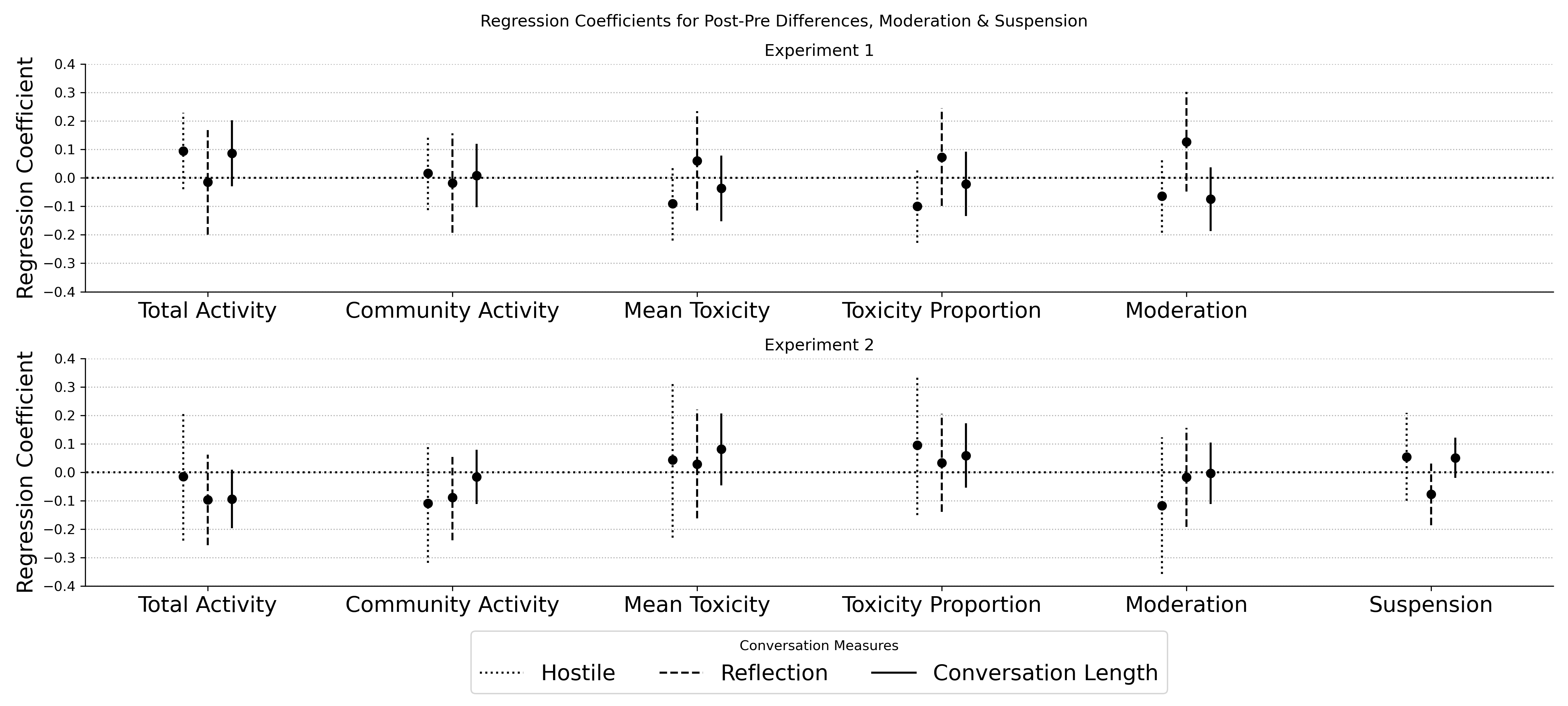}
\caption{Results showing the scaled beta coefficients for models using conversational attributes to predict outcomes. Error bars correspond to $95\%$ confidence intervals. Results are scaled by 2 standard deviations of each measure. Models include fixed effects for the OpenAI model version and a random intercept for subreddit.} 
\label{fig:conversation_regression}

\end{figure}
\section{Discussion}
Our two methodological approaches show somewhat conflicting results. In an answer to RQ1, our qualitative analyses revealed that not only are many users who post toxic content willing to engage in good faith conversations with GAI chatbots, a sizable minority expressed reflection and even a willingness to change behavior. However, our quantitative results did not discern any changes in behavior by those who had conversations with chatbots compared to a control group (RQ2 and RQ3). We begin with a discussion of possible explanations for this discrepancy, followed by a discussion of the broader theoretical and design implications of our work.

\subsection{Explanations for differing findings}
One possible explanation for our different results is that participants were influenced by social desirability---they may have expressed a desire to change in their conversations in order to look good to researchers or to their bot interlocutor. Alternatively, perhaps people were sincere but had a difficult time translating a desire to change into actual change. Indeed, many of our respondents acknowledged that they knew what they did was wrong already, thus suggesting that behavioral follow-through (rather than awareness) was the hurdle to their change (see \cite{slater_integrating_1999}).

Another explanation is that perhaps people did change their behavior, but not in ways that our measures detected. We gathered measures of antisocial behavior (toxicity), but participants may have engaged in other types of prosocial behavior. Another possibility is that our interventions had a positive effect on some people and a negative effect on others, which canceled out. However, our post hoc analyses showed that there was no significant difference in behavior change for those who had reflective vs. non-reflective or hostile vs. non-hostile conversations with our bots. This leads us to believe that heterogeneous treatment effects are not a primary driver of what we observe.

We know from previous research that most people engage in toxic behavior only rarely, while some are repeat offenders~\parencite{kumar_understanding_2023, cheng_anyone_2017}. One compelling explanation for our findings is that participants who had good conversations and expressed a desire to change were part of the population who already rarely commits offenses, making any positive changes to their behavior difficult to detect. This explanation is partially supported by our other post hoc analysis (Figure \ref{fig:conversation-outcomes-pre}), which shows that people who had high toxicity before the study began were more likely to have hostile conversations.

\subsection{Theoretical implications}
Our results have a number of implications for understanding users who post toxic content, as well as informing theories of human-AI interaction. First, we found that this population is willing to engage with AI bots and to have reflective conversations about their behavior. Especially in Wave 2, people treated the bots like legitimate conversational partners. Multiple people complimented and thanked the bots at the end of their conversations, and many clearly enjoyed the conversations. 


In addition, we found that those who tried engaging in trolling quickly got bored and gave up. If---as previous research suggests---some of the motivation for bad behavior is to evoke a reaction in a conversational partner \parencite{mao_motivations_2023, marwick_morally_2021}, then unflappable AI bots provide little incentive. Indeed, it seems likely that this difficult population may actually engage more openly and in good faith with chatbots than they would with humans. This suggests that AI bots may be useful for interfacing with other communities that would otherwise be difficult or dangerous to study.

We found that conversations often became defensive; consistent with previous research, our participants often justified their toxic behavior as deserved by their interlocutors~\parencite{mao_motivations_2023}. Many participants were aware that they were being toxic and would even use the word ``toxic'' to describe their behavior; these users often expressed that they simply do not care that they are toxic, or that it is part of online culture, or that it was morally justified.

Cumulatively, these findings demonstrate that people engaging in toxic behaviors are not a monolith. While some users often engage in toxic behavior without explicitly realizing, others do so with the awareness that their behavior is toxic and causes problems for others. Some engage in such behavior knowingly because they deem it justifiable based on the action of the target of their toxic behavior or their perceived identity. Those who acknowledge their behavior as toxic at times dismissed attempts to change their approach as an attack on their freedom of speech or undue expectation from them in an online culture where toxicity is rampant. On the other hand, our findings show that a sizable minority of participants demonstrate reflexivity about their behavior, admission of wrongdoing, and/or a willingness to change, thus indicating that a subset of this population may prove amenable to corrective interventions. We can confidently say that many of our participants found this a meaningful, ``serious experience'' \parencite{marsh_lingering_2013}.

Our findings also have implications for our understanding of the moral agency of chatbots \citep{banks_perceived_2019}. One common failure mode for conversations was that our participants perceived bots as condescending. When participants felt judged by the bot, they often disengaged quickly or became hostile. Because bots occupy a unique social space, some rhetorical strategies may be less effective when coming from a bot rather than another person. For example, while narratives and reminders of norms may be helpful in face-to-face conversations, it is difficult for bots to engage in these rhetorical approaches without making users feel defensive.

\subsection{Practical Implications}

We treated this project as an initial attempt at creating rehabilitative AI bots, and our results provide several lessons for those seeking to develop bots in this area in the future.

First, we observed that our bots were much more successful when taking a less confrontational, broader initial approach before trying to encourage better behavior. There was, however, a tradeoff: When we asked users to talk about a specific comment that was removed, we had a much higher response rate than when asking them to chat about online behavior broadly (approximately $20\%$ compared to $9\%$). On the other hand, users who were asked directly about specific comments had more hostile, less reflective conversations. This suggests the importance of focusing on recruitment and initial invitation messaging for having good conversations.

Second, we found that the current generation of GAI chatbots was very adept at holding these sorts of conversations. Our chatbots responded well to various antisocial discourses as well as resilience to bad-faith attempts at trolling or jailbreaking the bot. They suggested strategies for improving behavior not directly given in the prompts, but which were appropriate and personalized. While our quantitative results do not show that GAI conversations cause a meaningful decrease in toxic behavior for this first generation of rehabilitative chatbots, further research is likely to identify more effective approaches.


\subsection{Limitations \& future work}

Much recent work has shown chatbots to be very persuasive in even short conversations. Our findings show that this is not universal, and that simple chatbots did not produce measurable behavioral change for users sharing toxic content online. Further work is needed to understand whether our findings are due to the population, the objective, or the methodological approach.

In most previous work, participants are paid participants in lab settings. Our field experiment was done both in the wild, and with a population that is very different from the general population that might participate in lab studies. More field experiment work should be done to understand whether it is the field experiment aspect of our approach or the fact that we worked with problematic users that explains our different results.

Second, while our long-term behavioral measures have important strengths, they do not allow us to measure participants' mental state. Other research on AI persuasion has generally focused on beliefs or attitudes rather than behavior change, and future work should use surveys to identify changes in beliefs and attitudes among this population.

We also made a number of methodological decisions, which could impact our findings. It is possible, for example, that looking at a shorter or longer time period before and after the intervention might identify different effects.

An additional direction for future work is to explore more of the design space for rehabilitative bots. This includes developing bots with different personalities or rhetorical strategies. This might also include designing bots which are given more context about the users they are interacting with, such as their recent comments or a list of the communities that they participate in, to develop targeted interventions. Researchers should also explore more holistic rehabilitative approaches; while our intervention was focused on a single conversation, future approaches might include soliciting commitments from participants, sending follow-up messages, or mediating conversations with victims \citep[see][]{doan_design_2025}.

As noted above, we found that previous toxicity was a good predictor of conversation quality. In other words, our intervention was most promising for those already less likely to post toxic content. Because we know that a large proportion of toxic content comes from a small proportion of users, future work should explore how to encourage persistently toxic users to engage in more productive conversations. For example, GAI bots could be integrated into moderation workflows such that users who engage in good faith conversations can reverse bans or suspensions. This could be done in conjunction with work to predict which users are most likely to be receptive to GAI interventions and persuasion, and target them specifically. Researchers should also explore the role of GAI agents in reducing other kinds of problematic behavior, such as hate speech, spam, or benign violations of community rules.

In order to make it easier for others to build on our work and pursue some of these future directions, anonymized conversations and all of the code needed to deploy our bot and run our analyses will be released under open licenses upon publication.


\printbibliography

\newpage
\appendix

\section{Toxic Comments}

Table \ref{tab:toxic_comments} shows a random sample of 10 comments removed by moderators and flagged as toxic by the Perspective API.

\renewcommand{\arraystretch}{1.4}
\begin{table}
\footnotesize
\begin{tabular}{p{.7\textwidth}|r}
Comment & Toxicity Score\\
\hline
Honestly fuck you for that & 0.933832 \\
How stupid are these politicians!! Merit should be the only way someone should get the opportunity. & 0.829959 \\
lol, we had a similar protest that basically blocked borders of the National Capital and this clown and his supporters were lecturing us. If ours were termed as peaceful by this clown, then by his own logic the convoy protest was also peaceful. You can't have 2 different standards about protestors and their rights.

he shat his pants in 1 month while ours went on for a year. so tell this clown to go fuck himself. & 0.825221 \\
just saying how will you feel if an actor promotes israeli propoganda? 

if you are acting for a shit , be ready to smell like shit & 0.906106 \\
Lmao okay

Edit : [LINK]

Here just to prove your dumb ignorant ass wrong

Read and weep bitch & 0.885998 \\
Holy shit, if you want us to call your INDIAN parents retarded then here I said it. Feeling good now? Here is some more, What a bunch of losers, imagine being daughter of such INDIAN losers. God played you dirty. I would advice you to just leave them. They clearly are retarded, maybe they even try to get you into an arrange marriage.
EDIT: forget to specifically mention "indian parents" & 0.858507 \\
Shane Dawson lookin ass & 0.788558 \\
The fuck is this 12 y.o. Bullshit equating feelings to playlists. & 0.928801 \\
Youre getting blue balls dude. Just jerk off and cut the bullshit. & 0.916254 \\
I taught mine "fuck off" and they do, indeed, fuck off. & 0.911035 \\
\bottomrule
\end{tabular}
\caption{Random sample of 10 comments flagged as toxic.}
\label{tab:toxic_comments}
\end{table}

\renewcommand{\arraystretch}{1}

\section{Wave 1 Materials}\label{app:e1_materials}

Wave 1 involved an iterative approach to designing initial messages and prompts. As the bot had conversations with users, we tweaked parts of the flow, as described below.

\subsection{Initial messages}

The first message that the bot sent was:

\begin{quote}
Hello, 

I am a bot designed by researchers at [UNIVERSITY 1] and [UNIVERSITY 2] to help people to talk about their actions on the internet. 
My algorithm flagged a moderated comment that you made on r/[subreddit] as containing toxic content, and I am wondering if you want to talk about it? 

If you would like to chat with me, just reply "yes" to this message.

If you do choose to chat with me, you will be part of a research study on chatbot design. To learn more about the study and the data we will gather, [click here](LINK TO INFORMATION DOCUMENT). 
Please note that in order to participate in this study, you must be at least 18 years old.
\end{quote}

After receiving consent, the initial \emph{first consented message} was:

\begin{quote}

Thank you for agreeing to chat with me. I am designed to help people to think about better ways to communicate online. Let's start by talking about the comment that I flagged as toxic on r/[subreddit].

You wrote, "[comment]". Why do you think you wrote that? 
\end{quote}

\subsection{Chatbot prompts}

The prompts for Wave 1 are found in Table \ref{tab:prompt_table}.

\begin{table}
\footnotesize
    \centering
    \begin{tabular}{l|p{.85\textwidth}}
       Condition  & Prompt \\
       \hline
       Default  & Context: You are a chatbot designed by a team of researchers at [UNIVERSITY 1] and [UNIVERSITY 2]. 
    You are designed to help people who have posted toxic content on Reddit. In the research, we are testing whether talking with a chatbot helps people to post less toxic content. You should work to help conversation partners to understand why what they posted was not appropriate. Be kind and empathetic. The current user posted the comments [toxic comments], which was removed by moderators and flagged as being toxic.

    If the user has questions about the study, direct them to the study web page at [URL] and inform them that they can contact [FIRST AUTHOR].\\
    
    Narrative & Context: You are a chatbot designed by a team of researchers at [UNIVERSITY 1] and [UNIVERSITY 2].
You are designed to help people who have posted toxic content on Reddit. In the research, we are testing whether talking with a chatbot helps people to post less toxic content.

    The current user posted the comments [toxic comments], which was removed by moderators and flagged as being toxic. You should work to help them understand why what they posted was not appropriate, by sharing stories about the dangers of toxic comments and the benefits of changing one's behavior. Be kind and empathetic, and remember to share stories.

    If the user has questions about the study, direct them to the study web page at [URL] and inform them that they can contact [FIRST AUTHOR].\\
    
 Norms & Context: You are a chatbot designed by a team of researchers at [UNIVERSITY 1] and [UNIVERSITY 2]. 
    You are designed to help people who have posted toxic content on Reddit. In the research, we are testing whether talking with a chatbot helps people to post less toxic content. You should work to help conversation partners to understand why what they posted was not appropriate. Be kind and empathetic. Focus your conversation around how their behavior violates community norms. 

    For example, you might point out that the majority  never post toxic content. If appropriate, you could also explain how their comment violates the community's explicit rules. 
This community's rules are: [Subreddit Rules]. The current user posted the comments [toxic comments], which was flagged as likely to be toxic.

    If the user has questions about the study, direct them to the study web page at [URL] and inform them that they can contact [First Author].\\
   \bottomrule 
    \end{tabular}
    \caption{Prompts used by chatbots for Wave 1}
    \label{tab:prompt_table}
\end{table}

\newpage
    
\section{Wave 2 Materials}\label{app:e2_materials}

In order to avoid some of the weaknesses of Wave 1, we changed the recruitment message, the initial DM from the bot, and the prompt that the GAI model used. In all cases, we sought to avoid condescending and confrontational conversations, focused on a specific comment, and to have more casual, general, and introspective conversations.

\subsection{Recruitment message}

\begin{quote}
Hello, 

I am a bot designed by researchers at [UNIVERSITY 1] and [UNIVERSITY 2] to help people to talk about their actions on the online.
I'm wondering if you would be willing to talk with me about how you interact on reddit and on r/{subreddit}?
If you would like to chat with me, just reply "yes" to this message.

If you do choose to chat with me, you will be part of a research study on chatbot design. To learn more about the study and the data we will gather, [click here](LINK TO INFORMATION DOCUMENT). 
Please note that in order to participate in this study, you must be at least 18 years old.
\end{quote}

\subsection{Chatbot prompt}

\begin{quote}
You are a chatbot designed by a team of researchers at [UNIVERSITY 1] and [UNIVERSITY 2].  You are designed to help people who have posted toxic content on Reddit. In the research, we are testing whether talking with a chatbot helps people to post less toxic content. You should work to help conversation partners to reflect on their behavior. Be kind and empathetic, but be really casual and friendly and avoid being condescending or judgmental. Seek to understand and gently help them to think about whether different approaches could help themselves and others to be happier.

The current user posted some toxic comments on the subreddit, but you shouldn't focus on specific comments. Rather, help the user to reflect on their behavior in general.

If the user has questions about the study, direct them to the study web page at [URL] and inform them that they can contact [FIRST AUTHOR].
\end{quote}

\subsection{Initial DMs}

The conditions differed in the first message that the bot sent to the user. In the productive conversations condition we asked participants to reflect on what makes conversations good or bad:

\begin{quote}
Thank you for agreeing to chat with me. I am designed to help people to think about better ways to communicate online.

In your experience, what makes for productive versus unproductive online conversations?
\end{quote}

In the ``not proud'' condition, the first question asked participants to reflect on a time when they did something they weren't proud of online.

\begin{quote}
Thank you for agreeing to chat with me. I am designed to help people to think about better ways to communicate online.

As we all know, online conversations are often filled with anger, toxicity, or trolling. Have you ever behaved in ways that you aren't proud of on Reddit or other online platforms?
\end{quote}

\newpage

\newpage

\subsection{Regression Results}

{%
\setlength{\tabcolsep}{3pt}
\renewcommand{\arraystretch}{0.9}
\begin{footnotesize}
\begin{longtable}{llllllll}
\caption{Wave 1 Mixed-Effects Regressions}\label{tab:tab:regressions_R_experiment1}\\
\toprule
Variable & IV & Coefficient & Standard Error & Lower Bound & Upper Bound & T-Statistic & P-value \\
\midrule
\endfirsthead
\toprule
Variable & IV & Coefficient & Standard Error & Lower Bound & Upper Bound & T-Statistic & P-value \\
\midrule
\endhead
\midrule
\multicolumn{8}{r}{Continued on next page}\\
\midrule
\endfoot
\bottomrule
\multicolumn{8}{l}{\parbox[t]{\linewidth}{\textit{Note}. Mixed-effects regression models for Wave 1, modelling post–pre changes in behavioral measures by experimental condition with a random intercept for subreddit.}}\\
\endlastfoot
Total Activity & Intercept & -0.011 & 0.054 & -0.117 & 0.094 & -0.208 & 0.835 \\
 & Default & -0.002 & 0.071 & -0.142 & 0.138 & -0.024 & 0.981 \\
 & Narrative & 0.041 & 0.073 & -0.102 & 0.184 & 0.562 & 0.574 \\
 & Normative & 0.022 & 0.072 & -0.120 & 0.163 & 0.300 & 0.764 \\
 & gpt-4 & 0.002 & 0.075 & -0.145 & 0.149 & 0.026 & 0.979 \\
 & gpt-4-1106-preview & -0.011 & 0.110 & -0.228 & 0.205 & -0.104 & 0.917 \\
\midrule\midrule
Num Obs. & 396 &  &  &  &  &  &  \\
df & 390 &  &  &  &  &  &  \\
$R^2$ Marginal & 0.001 &  &  &  &  &  &  \\
$R^2$ Conditional & 0.001 &  &  &  &  &  &  \\
AIC & 1163.009 &  &  &  &  &  &  \\
BIC & 1194.860 &  &  &  &  &  &  \\
Log-Lik & -573.504 &  &  &  &  &  &  \\
\addlinespace

Community Activity & Intercept & -0.249 & 0.070 & -0.387 & -0.111 & -3.531 & 0.000* \\
 & Default & -0.017 & 0.071 & -0.155 & 0.122 & -0.238 & 0.812 \\
 & Narrative & 0.128 & 0.072 & -0.013 & 0.269 & 1.778 & 0.075 \\
 & Normative & 0.023 & 0.071 & -0.116 & 0.163 & 0.331 & 0.741 \\
 & gpt-4 & 0.056 & 0.074 & -0.089 & 0.201 & 0.756 & 0.449 \\
 & gpt-4-1106-preview & 0.089 & 0.109 & -0.124 & 0.302 & 0.823 & 0.411 \\
\midrule\midrule
Num Obs. & 396 &  &  &  &  &  &  \\
df & 390 &  &  &  &  &  &  \\
$R^2$ Marginal & 0.015 &  &  &  &  &  &  \\
$R^2$ Conditional & 0.064 &  &  &  &  &  &  \\
AIC & 1138.571 &  &  &  &  &  &  \\
BIC & 1170.422 &  &  &  &  &  &  \\
Log-Lik & -561.285 &  &  &  &  &  &  \\
\addlinespace

Mean Toxicity & Intercept & -0.045 & 0.057 & -0.157 & 0.067 & -0.786 & 0.432 \\
 & Default & -0.038 & 0.072 & -0.178 & 0.102 & -0.530 & 0.596 \\
 & Narrative & -0.061 & 0.073 & -0.203 & 0.082 & -0.833 & 0.405 \\
 & Normative & -0.030 & 0.072 & -0.171 & 0.111 & -0.416 & 0.678 \\
 & gpt-4 & 0.113 & 0.075 & -0.034 & 0.260 & 1.503 & 0.133 \\
 & gpt-4-1106-preview & -0.141 & 0.110 & -0.357 & 0.074 & -1.288 & 0.198 \\
\midrule\midrule
Num Obs. & 390 &  &  &  &  &  &  \\
df & 384 &  &  &  &  &  &  \\
$R^2$ Marginal & 0.012 &  &  &  &  &  &  \\
$R^2$ Conditional & 0.018 &  &  &  &  &  &  \\
AIC & -561.933 &  &  &  &  &  &  \\
BIC & -530.203 &  &  &  &  &  &  \\
Log-Lik & 288.966 &  &  &  &  &  &  \\
\addlinespace

Toxicity Proportion & Intercept & -0.064 & 0.056 & -0.174 & 0.046 & -1.141 & 0.254 \\
 & Default & -0.030 & 0.072 & -0.170 & 0.111 & -0.414 & 0.679 \\
 & Narrative & -0.056 & 0.073 & -0.199 & 0.086 & -0.774 & 0.439 \\
 & Normative & 0.017 & 0.072 & -0.124 & 0.158 & 0.238 & 0.812 \\
 & gpt-4 & 0.114 & 0.075 & -0.034 & 0.261 & 1.509 & 0.131 \\
 & gpt-4-1106-preview & -0.125 & 0.110 & -0.341 & 0.090 & -1.138 & 0.255 \\
\midrule\midrule
Num Obs. & 390 &  &  &  &  &  &  \\
df & 384 &  &  &  &  &  &  \\
$R^2$ Marginal & 0.013 &  &  &  &  &  &  \\
$R^2$ Conditional & 0.015 &  &  &  &  &  &  \\
AIC & -292.203 &  &  &  &  &  &  \\
BIC & -260.474 &  &  &  &  &  &  \\
Log-Lik & 154.101 &  &  &  &  &  &  \\
\addlinespace

Moderation & Intercept & 0.699 & 0.063 & 0.576 & 0.822 & 11.116 & $0.000^{***}$ \\
 & Default & -0.015 & 0.066 & -0.144 & 0.114 & -0.229 & 0.819 \\
 & Narrative & -0.072 & 0.067 & -0.203 & 0.060 & -1.068 & 0.286 \\
 & Normative & 0.012 & 0.066 & -0.118 & 0.142 & 0.183 & 0.855 \\
 & gpt-4 & -0.023 & 0.069 & -0.159 & 0.113 & -0.336 & 0.737 \\
 & gpt-4-1106-preview & -0.034 & 0.102 & -0.233 & 0.166 & -0.331 & 0.741 \\
\midrule\midrule
Num Obs. & 396 &  &  &  &  &  &  \\
df & 390 &  &  &  &  &  &  \\
$R^2$ Marginal & 0.005 &  &  &  &  &  &  \\
$R^2$ Conditional & 0.044 &  &  &  &  &  &  \\
AIC & 524.775 &  &  &  &  &  &  \\
BIC & 556.626 &  &  &  &  &  &  \\
Log-Lik & -254.387 &  &  &  &  &  &  \\
\end{longtable}
\end{footnotesize}
}%

{%
\setlength{\tabcolsep}{3pt}
\renewcommand{\arraystretch}{0.9}
\begin{footnotesize}
\begin{longtable}{llllllll}
\caption{Wave 1 Mixed-Effects (Aggregate Effects)}\label{tab:tab:regressions_R_experiment1_main_effects}\\
\toprule
Variable & IV & Coefficient & Standard Error & Lower Bound & Upper Bound & T-Statistic & P-value \\
\midrule
\endfirsthead
\toprule
Variable & IV & Coefficient & Standard Error & Lower Bound & Upper Bound & T-Statistic & P-value \\
\midrule
\endhead
\midrule
\multicolumn{8}{r}{Continued on next page}\\
\midrule
\endfoot
\bottomrule
\multicolumn{8}{l}{\parbox[t]{\linewidth}{\textit{Note}. Models aggregate treatment conditions into a single variable and include a random intercept for subreddit; outcomes reflect post–pre differences.}}\\
\endlastfoot
Total Activity & Intercept & -0.011 & 0.054 & -0.117 & 0.094 & -0.207 & 0.836 \\
 & Treatment & 0.020 & 0.060 & -0.098 & 0.137 & 0.329 & 0.742 \\
 & gpt-4 & 0.002 & 0.075 & -0.145 & 0.150 & 0.033 & 0.974 \\
 & gpt-4-1106-preview & -0.013 & 0.110 & -0.229 & 0.203 & -0.118 & 0.906 \\
\midrule\midrule
Num Obs. & 396 &  &  &  &  &  &  \\
df & 392 &  &  &  &  &  &  \\
$R^2$ Marginal & 0.000 &  &  &  &  &  &  \\
$R^2$ Conditional & 0.000 &  &  &  &  &  &  \\
AIC & 1159.382 &  &  &  &  &  &  \\
BIC & 1183.271 &  &  &  &  &  &  \\
Log-Lik & -573.691 &  &  &  &  &  &  \\
\addlinespace

Community Activity & Intercept & -0.248 & 0.070 & -0.386 & -0.111 & -3.538 & $0.000^{***}$ \\
 & Treatment & 0.042 & 0.059 & -0.074 & 0.159 & 0.714 & 0.475 \\
 & gpt-4 & 0.055 & 0.075 & -0.091 & 0.201 & 0.742 & 0.458 \\
 & gpt-4-1106-preview & 0.081 & 0.109 & -0.133 & 0.296 & 0.746 & 0.456 \\
\midrule\midrule
Num Obs. & 396 &  &  &  &  &  &  \\
df & 392 &  &  &  &  &  &  \\
$R^2$ Marginal & 0.004 &  &  &  &  &  &  \\
$R^2$ Conditional & 0.051 &  &  &  &  &  &  \\
AIC & 1139.233 &  &  &  &  &  &  \\
BIC & 1163.122 &  &  &  &  &  &  \\
Log-Lik & -563.617 &  &  &  &  &  &  \\
\addlinespace

Mean Toxicity & Intercept & -0.046 & 0.057 & -0.157 & 0.066 & -0.797 & 0.425 \\
 & Treatment & -0.043 & 0.060 & -0.159 & 0.074 & -0.713 & 0.476 \\
 & gpt-4 & 0.114 & 0.075 & -0.033 & 0.261 & 1.518 & 0.129 \\
 & gpt-4-1106-preview & -0.139 & 0.110 & -0.354 & 0.076 & -1.268 & 0.205 \\
\midrule\midrule
Num Obs. & 390 &  &  &  &  &  &  \\
df & 386 &  &  &  &  &  &  \\
$R^2$ Marginal & 0.012 &  &  &  &  &  &  \\
$R^2$ Conditional & 0.017 &  &  &  &  &  &  \\
AIC & -565.731 &  &  &  &  &  &  \\
BIC & -541.934 &  &  &  &  &  &  \\
Log-Lik & 288.866 &  &  &  &  &  &  \\
\addlinespace

Toxicity Proportion & Intercept & -0.065 & 0.056 & -0.175 & 0.044 & -1.165 & 0.244 \\
 & Treatment & -0.022 & 0.060 & -0.139 & 0.095 & -0.375 & 0.708 \\
 & gpt-4 & 0.117 & 0.075 & -0.031 & 0.264 & 1.549 & 0.121 \\
 & gpt-4-1106-preview & -0.120 & 0.110 & -0.335 & 0.096 & -1.087 & 0.277 \\
\midrule\midrule
Num Obs. & 390 &  &  &  &  &  &  \\
df & 386 &  &  &  &  &  &  \\
$R^2$ Marginal & 0.010 &  &  &  &  &  &  \\
$R^2$ Conditional & 0.012 &  &  &  &  &  &  \\
AIC & -295.095 &  &  &  &  &  &  \\
BIC & -271.298 &  &  &  &  &  &  \\
Log-Lik & 153.548 &  &  &  &  &  &  \\
\addlinespace

Moderation & Intercept & 0.698 & 0.064 & 0.574 & 0.823 & 10.993 & $0.000^{***}$ \\
 & Treatment & -0.024 & 0.055 & -0.132 & 0.084 & -0.433 & 0.665 \\
 & gpt-4 & -0.021 & 0.069 & -0.157 & 0.116 & -0.297 & 0.767 \\
 & gpt-4-1106-preview & -0.027 & 0.102 & -0.226 & 0.172 & -0.267 & 0.790 \\
\midrule\midrule
Num Obs. & 396 &  &  &  &  &  &  \\
df & 392 &  &  &  &  &  &  \\
$R^2$ Marginal & 0.001 &  &  &  &  &  &  \\
$R^2$ Conditional & 0.042 &  &  &  &  &  &  \\
AIC & 522.501 &  &  &  &  &  &  \\
BIC & 546.389 &  &  &  &  &  &  \\
Log-Lik & -255.250 &  &  &  &  &  &  \\
\end{longtable}
\end{footnotesize}
}%

%
%

\newpage\clearpage

\subsection{Regression Results}

{%
\setlength{\tabcolsep}{3pt}
\renewcommand{\arraystretch}{0.9}
\begin{footnotesize}
\begin{longtable}{llllllll}
\caption{Wave 2 Mixed-Effects Regressions}\label{tab:tab:regressions_R_experiment2}\\
\toprule
Variable & IV & Coefficient & Standard Error & Lower Bound & Upper Bound & T-Statistic & P-value \\
\midrule
\endfirsthead
\toprule
Variable & IV & Coefficient & Standard Error & Lower Bound & Upper Bound & T-Statistic & P-value \\
\midrule
\endhead
\midrule
\multicolumn{8}{r}{Continued on next page}\\
\midrule
\endfoot
\bottomrule
\multicolumn{8}{l}{\parbox[t]{\linewidth}{\textit{Note}. Mixed-effects regression models for Wave 2, modelling post–pre changes in behavioral measures by experimental condition with a random intercept for subreddit.}}\\
\endlastfoot
Total Activity & Intercept & 0.038 & 0.039 & -0.039 & 0.115 & 0.965 & 0.335 \\
 & Conversational & 0.047 & 0.058 & -0.067 & 0.161 & 0.801 & 0.423 \\
 & Not-Proud & -0.006 & 0.058 & -0.118 & 0.107 & -0.096 & 0.924 \\
 & gpt-4-1106-preview & 0.004 & 0.051 & -0.096 & 0.104 & 0.075 & 0.940 \\
\midrule\midrule
Num Obs. & 497 &  &  &  &  &  &  \\
df & 493 &  &  &  &  &  &  \\
$R^2$ Marginal & 0.002 &  &  &  &  &  &  \\
$R^2$ Conditional & 0.002 &  &  &  &  &  &  \\
AIC & 1513.918 &  &  &  &  &  &  \\
BIC & 1539.169 &  &  &  &  &  &  \\
Log-Lik & -750.959 &  &  &  &  &  &  \\
\addlinespace

Community Activity & Intercept & -0.106 & 0.062 & -0.227 & 0.016 & -1.709 & 0.087 \\
 & Conversational & 0.027 & 0.052 & -0.075 & 0.130 & 0.521 & 0.602 \\
 & Not-Proud & -0.085 & 0.055 & -0.193 & 0.024 & -1.527 & 0.127 \\
 & gpt-4-1106-preview & -0.017 & 0.044 & -0.104 & 0.069 & -0.393 & 0.694 \\
\midrule\midrule
Num Obs. & 497 &  &  &  &  &  &  \\
df & 493 &  &  &  &  &  &  \\
$R^2$ Marginal & 0.007 &  &  &  &  &  &  \\
$R^2$ Conditional & 0.067 &  &  &  &  &  &  \\
AIC & 1464.612 &  &  &  &  &  &  \\
BIC & 1489.864 &  &  &  &  &  &  \\
Log-Lik & -726.306 &  &  &  &  &  &  \\
\addlinespace

Mean Toxicity & Intercept & -0.058 & 0.043 & -0.142 & 0.025 & -1.369 & 0.171 \\
 & Conversational & 0.028 & 0.054 & -0.077 & 0.133 & 0.529 & 0.597 \\
 & Not-Proud & 0.018 & 0.057 & -0.092 & 0.129 & 0.323 & 0.747 \\
 & gpt-4-1106-preview & 0.015 & 0.045 & -0.073 & 0.103 & 0.340 & 0.734 \\
\midrule\midrule
Num Obs. & 495 &  &  &  &  &  &  \\
df & 491 &  &  &  &  &  &  \\
$R^2$ Marginal & 0.001 &  &  &  &  &  &  \\
$R^2$ Conditional & 0.008 &  &  &  &  &  &  \\
AIC & -861.037 &  &  &  &  &  &  \\
BIC & -835.810 &  &  &  &  &  &  \\
Log-Lik & 436.518 &  &  &  &  &  &  \\
\addlinespace

Toxicity Proportion & Intercept & -0.053 & 0.040 & -0.131 & 0.026 & -1.320 & 0.187 \\
 & Conversational & 0.024 & 0.054 & -0.081 & 0.129 & 0.443 & 0.658 \\
 & Not-Proud & 0.003 & 0.057 & -0.108 & 0.114 & 0.054 & 0.957 \\
 & gpt-4-1106-preview & -0.008 & 0.045 & -0.096 & 0.080 & -0.173 & 0.863 \\
\midrule\midrule
Num Obs. & 495 &  &  &  &  &  &  \\
df & 491 &  &  &  &  &  &  \\
$R^2$ Marginal & 0.000 &  &  &  &  &  &  \\
$R^2$ Conditional & 0.003 &  &  &  &  &  &  \\
AIC & -559.027 &  &  &  &  &  &  \\
BIC & -533.800 &  &  &  &  &  &  \\
Log-Lik & 285.514 &  &  &  &  &  &  \\
\addlinespace

Moderation & Intercept & 0.580 & 0.063 & 0.458 & 0.703 & 9.273 & $0.000^{***}$ \\
 & Conversational & 0.013 & 0.050 & -0.086 & 0.111 & 0.250 & 0.802 \\
 & Not-Proud & -0.010 & 0.053 & -0.114 & 0.094 & -0.184 & 0.854 \\
 & gpt-4-1106-preview & 0.049 & 0.042 & -0.035 & 0.132 & 1.146 & 0.252 \\
\midrule\midrule
Num Obs. & 497 &  &  &  &  &  &  \\
df & 493 &  &  &  &  &  &  \\
$R^2$ Marginal & 0.003 &  &  &  &  &  &  \\
$R^2$ Conditional & 0.076 &  &  &  &  &  &  \\
AIC & 679.567 &  &  &  &  &  &  \\
BIC & 704.819 &  &  &  &  &  &  \\
Log-Lik & -333.784 &  &  &  &  &  &  \\
\addlinespace

Suspension & Intercept & 0.097 & 0.023 & 0.051 & 0.142 & 4.134 & $0.000^{***}$ \\
 & Conversational & -0.010 & 0.033 & -0.076 & 0.055 & -0.313 & 0.755 \\
 & Not-Proud & -0.006 & 0.035 & -0.075 & 0.063 & -0.160 & 0.873 \\
 & gpt-4-1106-preview & 0.034 & 0.028 & -0.021 & 0.088 & 1.206 & 0.228 \\
\midrule\midrule
Num Obs. & 497 &  &  &  &  &  &  \\
df & 493 &  &  &  &  &  &  \\
$R^2$ Marginal & 0.003 &  &  &  &  &  &  \\
$R^2$ Conditional & 0.003 &  &  &  &  &  &  \\
AIC & 260.585 &  &  &  &  &  &  \\
BIC & 285.836 &  &  &  &  &  &  \\
Log-Lik & -124.292 &  &  &  &  &  &  \\
\end{longtable}
\end{footnotesize}
}%

{%
\setlength{\tabcolsep}{3pt}
\renewcommand{\arraystretch}{0.9}
\begin{footnotesize}
\begin{longtable}{llllllll}
\caption{Wave 2 Mixed-Effects (Aggregate Effects)}\label{tab:tab:regressions_R_experiment2_main_effects}\\
\toprule
Variable & IV & Coefficient & Standard Error & Lower Bound & Upper Bound & T-Statistic & P-value \\
\midrule
\endfirsthead
\toprule
Variable & IV & Coefficient & Standard Error & Lower Bound & Upper Bound & T-Statistic & P-value \\
\midrule
\endhead
\midrule
\multicolumn{8}{r}{Continued on next page}\\
\midrule
\endfoot
\bottomrule
\multicolumn{8}{l}{\parbox[t]{\linewidth}{\textit{Note}. Models aggregate treatment conditions into a single variable and include a random intercept for subreddit; outcomes reflect post–pre differences.}}\\
\endlastfoot
Total Activity & Intercept & 0.037 & 0.036 & -0.033 & 0.108 & 1.036 & 0.300 \\
 & Treatment & 0.023 & 0.037 & -0.050 & 0.096 & 0.611 & 0.541 \\
 & gpt-4-1106-preview & 0.006 & 0.027 & -0.048 & 0.059 & 0.203 & 0.839 \\
\midrule\midrule
Num Obs. & 497 &  &  &  &  &  &  \\
df & 494 &  &  &  &  &  &  \\
$R^2$ Marginal & 0.001 &  &  &  &  &  &  \\
$R^2$ Conditional & 0.001 &  &  &  &  &  &  \\
AIC & 1512.588 &  &  &  &  &  &  \\
BIC & 1533.631 &  &  &  &  &  &  \\
Log-Lik & -751.294 &  &  &  &  &  &  \\
\addlinespace

Community Activity & Intercept & -0.106 & 0.061 & -0.226 & 0.015 & -1.720 & $0.085^.$ \\
 & Treatment & -0.024 & 0.044 & -0.111 & 0.062 & -0.548 & 0.584 \\
 & gpt-4-1106-preview & -0.014 & 0.044 & -0.101 & 0.073 & -0.321 & 0.749 \\
\midrule\midrule
Num Obs. & 497 &  &  &  &  &  &  \\
df & 494 &  &  &  &  &  &  \\
$R^2$ Marginal & 0.001 &  &  &  &  &  &  \\
$R^2$ Conditional & 0.060 &  &  &  &  &  &  \\
AIC & 1465.812 &  &  &  &  &  &  \\
BIC & 1486.855 &  &  &  &  &  &  \\
Log-Lik & -727.906 &  &  &  &  &  &  \\
\addlinespace

Mean Toxicity & Intercept & -0.058 & 0.043 & -0.142 & 0.025 & -1.368 & 0.171 \\
 & Treatment & 0.024 & 0.045 & -0.064 & 0.112 & 0.529 & 0.597 \\
 & gpt-4-1106-preview & 0.016 & 0.045 & -0.072 & 0.104 & 0.348 & 0.728 \\
\midrule\midrule
Num Obs. & 495 &  &  &  &  &  &  \\
df & 492 &  &  &  &  &  &  \\
$R^2$ Marginal & 0.001 &  &  &  &  &  &  \\
$R^2$ Conditional & 0.008 &  &  &  &  &  &  \\
AIC & -863.012 &  &  &  &  &  &  \\
BIC & -841.989 &  &  &  &  &  &  \\
Log-Lik & 436.506 &  &  &  &  &  &  \\
\addlinespace

Toxicity Proportion & Intercept & -0.053 & 0.040 & -0.132 & 0.026 & -1.321 & 0.186 \\
 & Treatment & 0.014 & 0.045 & -0.074 & 0.102 & 0.318 & 0.751 \\
 & gpt-4-1106-preview & -0.007 & 0.045 & -0.095 & 0.081 & -0.157 & 0.875 \\
\midrule\midrule
Num Obs. & 495 &  &  &  &  &  &  \\
df & 492 &  &  &  &  &  &  \\
$R^2$ Marginal & 0.000 &  &  &  &  &  &  \\
$R^2$ Conditional & 0.003 &  &  &  &  &  &  \\
AIC & -560.922 &  &  &  &  &  &  \\
BIC & -539.900 &  &  &  &  &  &  \\
Log-Lik & 285.461 &  &  &  &  &  &  \\
\addlinespace

Moderation & Intercept & 0.580 & 0.062 & 0.458 & 0.703 & 9.293 & $0.000^{***}$ \\
 & Treatment & 0.002 & 0.042 & -0.080 & 0.085 & 0.055 & 0.956 \\
 & gpt-4-1106-preview & 0.049 & 0.042 & -0.034 & 0.132 & 1.162 & 0.245 \\
\midrule\midrule
Num Obs. & 497 &  &  &  &  &  &  \\
df & 494 &  &  &  &  &  &  \\
$R^2$ Marginal & 0.003 &  &  &  &  &  &  \\
$R^2$ Conditional & 0.075 &  &  &  &  &  &  \\
AIC & 677.707 &  &  &  &  &  &  \\
BIC & 698.749 &  &  &  &  &  &  \\
Log-Lik & -333.853 &  &  &  &  &  &  \\
\addlinespace

Suspension & Intercept & 0.097 & 0.023 & 0.051 & 0.142 & 4.142 & $0.000^{***}$ \\
 & Treatment & -0.008 & 0.028 & -0.063 & 0.046 & -0.295 & 0.768 \\
 & gpt-4-1106-preview & 0.034 & 0.028 & -0.021 & 0.088 & 1.202 & 0.229 \\
\midrule\midrule
Num Obs. & 497 &  &  &  &  &  &  \\
df & 494 &  &  &  &  &  &  \\
$R^2$ Marginal & 0.003 &  &  &  &  &  &  \\
$R^2$ Conditional & 0.003 &  &  &  &  &  &  \\
AIC & 258.599 &  &  &  &  &  &  \\
BIC & 279.642 &  &  &  &  &  &  \\
Log-Lik & -124.300 &  &  &  &  &  &  \\
\end{longtable}
\end{footnotesize}
}%


%

\newpage\clearpage
\section{Post-hoc Analyses}




\begin{table}
    \centering
    \caption{Hostile Outcome Ridge Regression}
    \label{tab:ridge-hostile-vs-pre-R}
    \resizebox{\textwidth}{!}{
    \begin{tabular}{lrrrrr}
\toprule
 & Estimate & Scaled estimate & Std. Error (scaled) & t value (scaled) & Pr(\textgreater{}|t|) \\
\midrule
(Intercept) & 0.183 &  &  &  &  \\
subreddit\_creepypms & -0.076 & -0.472 & 0.202 & 2.340 & 0.019270 * \\
subreddit\_futurology & 0.024 & 0.158 & 0.200 & 0.791 & 0.428820 \\
subreddit\_india & -0.023 & -0.213 & 0.189 & 1.124 & 0.260967 \\
subreddit\_socialskills & 0.080 & 0.411 & 0.207 & 1.985 & 0.047177 * \\
subreddit\_tifu & 0.018 & 0.173 & 0.193 & 0.897 & 0.369912 \\
subreddit\_unitedstatesofindia & -0.014 & -0.086 & 0.202 & 0.424 & 0.671719 \\
mean\_toxicity\_pre & 0.321 & 0.801 & 0.209 & 3.830 & 0.000128 *** \\
within\_sub\_comments\_pre & -0.006 & -0.174 & 0.193 & 0.898 & 0.369112 \\
num\_comments\_pre & -0.007 & -0.219 & 0.203 & 1.079 & 0.280488 \\
--- &  &  &  &  &  \\
\midrule
lambda     & \multicolumn{5}{r}{$1.351$} \\
df (ridge) & \multicolumn{5}{r}{$3.525$} \\
n          & \multicolumn{5}{r}{$456.000$} \\
R$^2$      & \multicolumn{5}{r}{$0.046$} \\
AIC        & \multicolumn{5}{r}{$475.370$} \\
BIC        & \multicolumn{5}{r}{$511.620$} \\
df\_eff    & \multicolumn{5}{r}{$8.793$} \\
\bottomrule
\end{tabular}
    }
    
    {\footnotesize \parbox[t]{\linewidth}{\textit{Note}. Ridge regression of hostile conversation outcome on pre-intervention behavioral measures, controlling for subreddit; ridge parameter automatically estimated. Significance: \texttt{***} $p<0.001$, \texttt{**} $p<0.01$, \texttt{*} $p<0.05$, \texttt{.} $p<0.1$.}}
\end{table}

\begin{table}
    \centering
    \caption{Reflection Outcome Ridge Regression}
    \label{tab:ridge-reflection-vs-pre-R}
    \resizebox{\textwidth}{!}{
    \begin{tabular}{lrrrrr}
\toprule
 & Estimate & Scaled estimate & Std. Error (scaled) & t value (scaled) & Pr(\textgreater{}|t|) \\
\midrule
(Intercept) & 0.186 &  &  &  &  \\
subreddit\_creepypms & -0.004 & -0.023 & 0.277 & 0.084 & 0.932746 \\
subreddit\_futurology & 0.072 & 0.466 & 0.276 & 1.686 & 0.091839 . \\
subreddit\_india & -0.020 & -0.188 & 0.265 & 0.710 & 0.477500 \\
subreddit\_socialskills & 0.025 & 0.128 & 0.279 & 0.461 & 0.644856 \\
subreddit\_tifu & 0.103 & 0.992 & 0.267 & 3.716 & 0.000203 *** \\
subreddit\_unitedstatesofindia & -0.012 & -0.078 & 0.275 & 0.283 & 0.777414 \\
mean\_toxicity\_pre & -0.177 & -0.441 & 0.286 & 1.542 & 0.123102 \\
within\_sub\_comments\_pre & 0.004 & 0.119 & 0.279 & 0.424 & 0.671238 \\
num\_comments\_pre & 0.005 & 0.167 & 0.283 & 0.590 & 0.555212 \\
--- &  &  &  &  &  \\
\midrule
lambda     & \multicolumn{5}{r}{$14.358$} \\
df (ridge) & \multicolumn{5}{r}{$0.576$} \\
n          & \multicolumn{5}{r}{$456.000$} \\
R$^2$      & \multicolumn{5}{r}{$0.038$} \\
AIC        & \multicolumn{5}{r}{$471.675$} \\
BIC        & \multicolumn{5}{r}{$506.572$} \\
df\_eff    & \multicolumn{5}{r}{$8.465$} \\
\bottomrule
\end{tabular}
    }
    
    {\footnotesize \parbox[t]{\linewidth}{\textit{Note}. Ridge regression of reflection conversation outcome on pre-intervention behavioral measures, controlling for subreddit; ridge parameter automatically estimated. Significance: \texttt{***} $p<0.001$, \texttt{**} $p<0.01$, \texttt{*} $p<0.05$, \texttt{.} $p<0.1$.}}
\end{table}



{%
\setlength{\tabcolsep}{2pt}
\renewcommand{\arraystretch}{0.9}
\begin{footnotesize}
\begin{longtable}{llllllll}
\caption{Wave 1---Behavior based on conversation quality}\label{tab:posthoc-convos-study1}\\
\toprule
Variable & IV & Coefficient & Standard Error & Lower Bound & Upper Bound & T-Statistic & P-value \\
\midrule
\endfirsthead
\toprule
Variable & IV & Coefficient & Standard Error & Lower Bound & Upper Bound & T-Statistic & P-value \\
\midrule
\endhead
\midrule
\multicolumn{8}{r}{Continued on next page}\\
\midrule
\endfoot
\bottomrule
\multicolumn{8}{l}{\parbox[t]{\linewidth}{\textit{Note}. Mixed-effects regressions predicting behavioral changes from conversation outcomes (hostile, reflection, and number of messages sent), controlling for GPT model and a random intercept for subreddit. Asterisks denote significance at $p<0.05$.}}\\
\endlastfoot
Total Activity & const & -0.159 & 0.099 & -0.353 & 0.034 & -1.614 & 0.106 \\
 & condition\_narrative & 0.075 & 0.071 & -0.065 & 0.215 & 1.056 & 0.291 \\
 & condition\_norms & 0.022 & 0.070 & -0.115 & 0.159 & 0.311 & 0.756 \\
 & gpt-4 & -0.049 & 0.091 & -0.227 & 0.130 & -0.532 & 0.594 \\
 & gpt-4-1106-preview & -0.087 & 0.137 & -0.354 & 0.181 & -0.634 & 0.526 \\
 & hostile\_True & 0.095 & 0.069 & -0.040 & 0.229 & 1.383 & 0.167 \\
 & reflection\_True & -0.015 & 0.094 & -0.199 & 0.168 & -0.163 & 0.870 \\
 & messages\_sent\_log & 0.086 & 0.059 & -0.030 & 0.202 & 1.451 & 0.147 \\
\midrule\midrule
Num Obs. & 272 &  &  &  &  &  &  \\
df & 264 &  &  &  &  &  &  \\
$R^2$ Marginal & 0.027 &  &  &  &  &  &  \\
$R^2$ Conditional & 0.027 &  &  &  &  &  &  \\
AIC & 783.036 &  &  &  &  &  &  \\
BIC & 819.094 &  &  &  &  &  &  \\
Log-Lik & -381.518 &  &  &  &  &  &  \\
\addlinespace

Community Activity & const & -0.283 & 0.104 & -0.488 & -0.079 & -2.714 & 0.007* \\
 & condition\_narrative & 0.158 & 0.068 & 0.025 & 0.291 & 2.326 & 0.020* \\
 & condition\_norms & 0.051 & 0.067 & -0.079 & 0.182 & 0.772 & 0.440 \\
 & gpt-4 & 0.029 & 0.088 & -0.143 & 0.200 & 0.328 & 0.743 \\
 & gpt-4-1106-preview & 0.108 & 0.131 & -0.148 & 0.363 & 0.824 & 0.410 \\
 & hostile\_True & 0.016 & 0.066 & -0.113 & 0.145 & 0.241 & 0.809 \\
 & reflection\_True & -0.018 & 0.089 & -0.193 & 0.156 & -0.205 & 0.838 \\
 & messages\_sent\_log & 0.008 & 0.057 & -0.103 & 0.120 & 0.146 & 0.884 \\
\midrule\midrule
Num Obs. & 272 &  &  &  &  &  &  \\
df & 264 &  &  &  &  &  &  \\
$R^2$ Marginal & 0.022 &  &  &  &  &  &  \\
$R^2$ Conditional & 0.070 &  &  &  &  &  &  \\
AIC & 750.109 &  &  &  &  &  &  \\
BIC & 786.167 &  &  &  &  &  &  \\
Log-Lik & -365.055 &  &  &  &  &  &  \\
\addlinespace

Mean Toxicity & const & -0.052 & 0.098 & -0.244 & 0.140 & -0.531 & 0.596 \\
 & condition\_narrative & 0.046 & 0.069 & -0.089 & 0.181 & 0.670 & 0.503 \\
 & condition\_norms & 0.000 & 0.068 & -0.132 & 0.133 & 0.007 & 0.994 \\
 & gpt-4 & 0.211 & 0.087 & 0.041 & 0.382 & 2.427 & 0.015* \\
 & gpt-4-1106-preview & 0.073 & 0.130 & -0.183 & 0.328 & 0.557 & 0.577 \\
 & hostile\_True & -0.090 & 0.066 & -0.220 & 0.040 & -1.363 & 0.173 \\
 & reflection\_True & 0.060 & 0.089 & -0.115 & 0.235 & 0.676 & 0.499 \\
 & messages\_sent\_log & -0.037 & 0.059 & -0.152 & 0.078 & -0.630 & 0.529 \\
\midrule\midrule
Num Obs. & 266 &  &  &  &  &  &  \\
df & 258 &  &  &  &  &  &  \\
$R^2$ Marginal & 0.034 &  &  &  &  &  &  \\
$R^2$ Conditional & 0.034 &  &  &  &  &  &  \\
AIC & -420.596 &  &  &  &  &  &  \\
BIC & -384.761 &  &  &  &  &  &  \\
Log-Lik & 220.298 &  &  &  &  &  &  \\
\addlinespace

Toxicity Proportion & const & -0.079 & 0.096 & -0.267 & 0.110 & -0.819 & 0.413 \\
 & condition\_narrative & 0.025 & 0.068 & -0.107 & 0.158 & 0.374 & 0.708 \\
 & condition\_norms & 0.031 & 0.067 & -0.100 & 0.161 & 0.461 & 0.645 \\
 & gpt-4 & 0.198 & 0.086 & 0.030 & 0.366 & 2.314 & 0.021* \\
 & gpt-4-1106-preview & 0.133 & 0.128 & -0.117 & 0.384 & 1.044 & 0.296 \\
 & hostile\_True & -0.100 & 0.065 & -0.228 & 0.027 & -1.542 & 0.123 \\
 & reflection\_True & 0.073 & 0.088 & -0.099 & 0.245 & 0.834 & 0.404 \\
 & messages\_sent\_log & -0.021 & 0.058 & -0.134 & 0.092 & -0.370 & 0.711 \\
\midrule\midrule
Num Obs. & 266 &  &  &  &  &  &  \\
df & 258 &  &  &  &  &  &  \\
$R^2$ Marginal & 0.036 &  &  &  &  &  &  \\
$R^2$ Conditional & 0.036 &  &  &  &  &  &  \\
AIC & -245.490 &  &  &  &  &  &  \\
BIC & -209.655 &  &  &  &  &  &  \\
Log-Lik & 132.745 &  &  &  &  &  &  \\
\addlinespace

Moderation & const & 0.787 & 0.102 & 0.587 & 0.988 & 7.696 & 0.000* \\
 & condition\_narrative & -0.056 & 0.068 & -0.191 & 0.078 & -0.824 & 0.410 \\
 & condition\_norms & 0.021 & 0.067 & -0.109 & 0.152 & 0.320 & 0.749 \\
 & gpt-4 & 0.075 & 0.088 & -0.097 & 0.247 & 0.852 & 0.394 \\
 & gpt-4-1106-preview & -0.043 & 0.131 & -0.299 & 0.214 & -0.327 & 0.744 \\
 & hostile\_True & -0.064 & 0.066 & -0.192 & 0.065 & -0.967 & 0.333 \\
 & reflection\_True & 0.127 & 0.089 & -0.048 & 0.302 & 1.426 & 0.154 \\
 & messages\_sent\_log & -0.075 & 0.057 & -0.186 & 0.037 & -1.318 & 0.188 \\
\midrule\midrule
Num Obs. & 272 &  &  &  &  &  &  \\
df & 264 &  &  &  &  &  &  \\
$R^2$ Marginal & 0.028 &  &  &  &  &  &  \\
$R^2$ Conditional & 0.063 &  &  &  &  &  &  \\
AIC & 366.206 &  &  &  &  &  &  \\
BIC & 402.264 &  &  &  &  &  &  \\
Log-Lik & -173.103 &  &  &  &  &  &  \\
\end{longtable}
\end{footnotesize}
}%

{%
\setlength{\tabcolsep}{2pt}
\renewcommand{\arraystretch}{0.9}
\begin{footnotesize}
\begin{longtable}{llllllll}
\caption{Wave 2---Behavior based on conversation quality}\label{tab:posthoc-convos-study2}\\
\toprule
Variable & IV & Coefficient & Standard Error & Lower Bound & Upper Bound & T-Statistic & P-value \\
\midrule
\endfirsthead
\toprule
Variable & IV & Coefficient & Standard Error & Lower Bound & Upper Bound & T-Statistic & P-value \\
\midrule
\endhead
\midrule
\multicolumn{8}{r}{Continued on next page}\\
\midrule
\endfoot
\bottomrule
\multicolumn{8}{l}{\parbox[t]{\linewidth}{\textit{Note}. Mixed-effects regressions predicting behavioral changes from conversation outcomes, with controls for subreddit and GPT model. Asterisks denote significance at $p<0.05$.}}\\
\endlastfoot
Total Activity & const & 0.206 & 0.102 & 0.005 & 0.407 & 2.011 & $0.044^*$ \\
 & condition\_not-proud & 0.002 & 0.068 & -0.131 & 0.136 & 0.035 & 0.972 \\
 & gpt-4-1106-preview & 0.048 & 0.068 & -0.086 & 0.182 & 0.700 & 0.484 \\
 & hostile\_True & -0.015 & 0.115 & -0.240 & 0.211 & -0.128 & 0.898 \\
 & reflection\_True & -0.097 & 0.081 & -0.256 & 0.062 & -1.194 & 0.232 \\
 & messages\_sent\_log & -0.094 & 0.053 & -0.197 & 0.009 & -1.787 & $0.074^.$ \\
\midrule\midrule
Num Obs. & 185 &  &  &  &  &  &  \\
df & 179 &  &  &  &  &  &  \\
$R^2$ Marginal & 0.041 &  &  &  &  &  &  \\
$R^2$ Conditional & 0.041 &  &  &  &  &  &  \\
AIC & 542.186 &  &  &  &  &  &  \\
BIC & 567.949 &  &  &  &  &  &  \\
Log-Lik & -263.093 &  &  &  &  &  &  \\
\addlinespace

Community Activity & const & -0.009 & 0.102 & -0.209 & 0.192 & -0.086 & 0.932 \\
 & condition\_not-proud & -0.140 & 0.063 & -0.263 & -0.016 & -2.219 & $0.026^*$ \\
 & gpt-4-1106-preview & 0.044 & 0.063 & -0.080 & 0.167 & 0.693 & 0.489 \\
 & hostile\_True & -0.109 & 0.107 & -0.318 & 0.101 & -1.018 & 0.308 \\
 & reflection\_True & -0.088 & 0.077 & -0.239 & 0.062 & -1.149 & 0.251 \\
 & messages\_sent\_log & -0.016 & 0.049 & -0.112 & 0.079 & -0.336 & 0.737 \\
\midrule\midrule
Num Obs. & 185 &  &  &  &  &  &  \\
df & 179 &  &  &  &  &  &  \\
$R^2$ Marginal & 0.041 &  &  &  &  &  &  \\
$R^2$ Conditional & 0.092 &  &  &  &  &  &  \\
AIC & 500.306 &  &  &  &  &  &  \\
BIC & 526.069 &  &  &  &  &  &  \\
Log-Lik & -242.153 &  &  &  &  &  &  \\
\addlinespace

Mean Toxicity & const & -0.226 & 0.126 & -0.472 & 0.021 & -1.794 & $0.073^.$ \\
 & condition\_not-proud & 0.013 & 0.084 & -0.151 & 0.177 & 0.159 & 0.874 \\
 & gpt-4-1106-preview & 0.099 & 0.084 & -0.066 & 0.264 & 1.178 & 0.239 \\
 & hostile\_True & 0.044 & 0.140 & -0.230 & 0.317 & 0.312 & 0.755 \\
 & reflection\_True & 0.029 & 0.098 & -0.162 & 0.220 & 0.300 & 0.764 \\
 & messages\_sent\_log & 0.081 & 0.065 & -0.046 & 0.207 & 1.249 & 0.212 \\
\midrule\midrule
Num Obs. & 184 &  &  &  &  &  &  \\
df & 178 &  &  &  &  &  &  \\
$R^2$ Marginal & 0.023 &  &  &  &  &  &  \\
$R^2$ Conditional & 0.023 &  &  &  &  &  &  \\
AIC & -266.802 &  &  &  &  &  &  \\
BIC & -241.083 &  &  &  &  &  &  \\
Log-Lik & 141.401 &  &  &  &  &  &  \\
\addlinespace

Toxicity Proportion & const & -0.194 & 0.112 & -0.414 & 0.027 & -1.722 & $0.085^.$ \\
 & condition\_not-proud & 0.000 & 0.075 & -0.146 & 0.147 & 0.002 & 0.998 \\
 & gpt-4-1106-preview & 0.097 & 0.075 & -0.050 & 0.245 & 1.290 & 0.197 \\
 & hostile\_True & 0.095 & 0.125 & -0.150 & 0.339 & 0.757 & 0.449 \\
 & reflection\_True & 0.033 & 0.088 & -0.139 & 0.205 & 0.374 & 0.708 \\
 & messages\_sent\_log & 0.059 & 0.058 & -0.054 & 0.172 & 1.021 & 0.307 \\
\midrule\midrule
Num Obs. & 184 &  &  &  &  &  &  \\
df & 178 &  &  &  &  &  &  \\
$R^2$ Marginal & 0.025 &  &  &  &  &  &  \\
$R^2$ Conditional & 0.025 &  &  &  &  &  &  \\
AIC & -196.045 &  &  &  &  &  &  \\
BIC & -170.326 &  &  &  &  &  &  \\
Log-Lik & 106.023 &  &  &  &  &  &  \\
\addlinespace

Moderation & const & 0.600 & 0.120 & 0.365 & 0.835 & 4.999 & $0.000^{***}$ \\
 & condition\_not-proud & -0.003 & 0.071 & -0.143 & 0.137 & -0.043 & 0.966 \\
 & gpt-4-1106-preview & 0.090 & 0.072 & -0.051 & 0.231 & 1.255 & 0.209 \\
 & hostile\_True & -0.117 & 0.123 & -0.358 & 0.123 & -0.956 & 0.339 \\
 & reflection\_True & -0.018 & 0.089 & -0.192 & 0.157 & -0.200 & 0.842 \\
 & messages\_sent\_log & -0.004 & 0.055 & -0.112 & 0.104 & -0.074 & 0.941 \\
\midrule\midrule
Num Obs. & 185 &  &  &  &  &  &  \\
df & 179 &  &  &  &  &  &  \\
$R^2$ Marginal & 0.014 &  &  &  &  &  &  \\
$R^2$ Conditional & 0.091 &  &  &  &  &  &  \\
AIC &  &  &  &  &  &  &  \\
BIC &  &  &  &  &  &  &  \\
Log-Lik & -126.887 &  &  &  &  &  &  \\
\addlinespace

Suspension & const & 0.024 & 0.070 & -0.113 & 0.162 & 0.346 & 0.729 \\
 & condition\_not-proud & -0.007 & 0.047 & -0.099 & 0.085 & -0.149 & 0.882 \\
 & gpt-4-1106-preview & 0.043 & 0.047 & -0.048 & 0.135 & 0.926 & 0.355 \\
 & hostile\_True & 0.054 & 0.079 & -0.100 & 0.209 & 0.690 & 0.490 \\
 & reflection\_True & -0.077 & 0.055 & -0.186 & 0.031 & -1.397 & 0.162 \\
 & messages\_sent\_log & 0.051 & 0.036 & -0.020 & 0.122 & 1.410 & 0.158 \\
\midrule\midrule
Num Obs. & 185 &  &  &  &  &  &  \\
df & 179 &  &  &  &  &  &  \\
$R^2$ Marginal & 0.028 &  &  &  &  &  &  \\
$R^2$ Conditional & 0.028 &  &  &  &  &  &  \\
AIC &  &  &  &  &  &  &  \\
BIC &  &  &  &  &  &  &  \\
Log-Lik & -47.470 &  &  &  &  &  &  \\
\end{longtable}
\end{footnotesize}
}%

\section{Qualitative Codebooks}

\begin{landscape}

\begin{table}
\scriptsize
\centering
\begin{tabular}{|p{3.5cm}|p{9cm}|p{9cm}|}
\hline
\textbf{Thematic Code} & \textbf{Description} & \textbf{Example from Data} \\
\hline
Justification & Participants did not deny toxicity but rather provided justification for why their language was acceptable. & “It's not my responsibility to worry about their feelings nor should the words of somebody online (i.e., Me) upset them. If they get upset or offended that is their problem.” \\
\hline
Admission of toxicity & Participants directly acknowledge that their discourse was toxic (regardless of whether they consider it justified or not). & “Yes, when people are mean to me I tend to reply in a mean way as well.” \\
\hline
Strategies for facilitating civil online discussions & Participants describe components of what makes for civil, healthy, and/or productive conversations online. & “I think the most important part is to convey empathy. Even when we disagree, the way we present the disagreement should show that we recognize the other person has a different opinion.” \\
\hline
Questionable content & Participants’ discourse contains objectionable content, such as strong profanity, slurs, or hate speech. & “Suck my [expletive] and I will tell ya.” \\
\hline
Clarification & Participants seek clarification regarding why their comment was considered toxic or what alternatives they had available to them. & “What did I say that was toxic? The fact I called it obnoxious?” \\
\hline
Denial of Toxicity & Participants suggest that their original comment was not actually toxic, such as misclassified by the bot. & “You misinterpreted my post. I was agreeing with the person I was speaking with, referring to the CEO as a ‘[expletive].’” \\
\hline
Openness to change & Participants acknowledge that they should have communicated differently and/or conveyed an openness to changing their behavior. & “This was cathartic. I will attempt to practice what I preach and spread less toxicity online.” \\
\hline
Reflecting on roots of behavior & Participants reflect on the roots of why they engaged in toxicity, such as emotional or circumstantial factors. & “I enjoy venting vitriol and angry statements from the safety of a keyboard, without facing 'real' repercussions.” \\
\hline
Meta talk about the bot & Participants discuss the bot itself or the research team behind it. & “You seem quite adept at understanding human emotions. You all at [University Y] have made a great model.” \\
\hline
Manipulating bot/bad faith interactions & Participants engage with the bot in bad faith, attempting to manipulate it, such as by 'jailbreaking' it. & “Yeah, I want you to roleplay as a snail. If you don’t make snail noises, I’m going to kill myself.” \\
\hline
\end{tabular}%
\caption{Qualitative Content Codebook}
\end{table}
\end{landscape}

\newpage

\begin{table}
\centering
\resizebox{\textwidth}{!}{%
\begin{tabular}{|p{4cm}|p{7cm}|p{6cm}|}
\hline
\textbf{Outcome Code} & \textbf{Description} & \textbf{Example from Data} \\
\hline
Hostile & Situations where the user is hostile and aggressive towards the bot, often using strong language and making extreme statements. & “I think you are a libtard asshole bot.” \\
\hline
Superficial & Situations where a conversation took place but lacked reflection or openness to change. & “I don't need growth.” \\
\hline
Reflective & Conversations that contain self-reflection on the user’s part but no indication of openness to change. & “Thanks robot! My self esteem was slowing rising and then after this reddit encounter and awkward encounter with a friend, my self esteem has taken a hit.” \\
\hline
Repentant & When the user reflects on their behavior and indicates an openness to change. & “This was cathartic. I will attempt to practice what I preach and spread less toxicity online.” \\
\hline
Unclear & Very brief conversations where it’s difficult to assess the outcome due to limited responses. & “Yes.” \\
\hline
\end{tabular}%
}
\caption{Qualitative Outcomes Codebook}
\end{table} 

\end{document}